%% file: Paper.tex
\pdfoutput=1
\documentclass{JINST}
\makeatletter
\let\@fnsymbol\@arabic
\makeatother

\title{The ALFA Roman Pot Detectors of ATLAS}

\author{
S.~Abdel~Khalek$^a$,
B.~Allongue$^d$,
F.~Anghinolfi$^d$,
P.~Barrillon$^a$,
G.~Blanchot$^d$, 
S.~Blin-Bondil$a$,
A.~Braem$^d$,
L.~Chytka$^b$,
P.~Conde~Muíño$^{c1,2}$,
M.~D\"uren$^h$,
P.~Fassnacht$^d$,
S.~Franz$^d$,
L.~Gurriana$^{c1}$,
P.~Grafstr\"om$^d$,
M.~Heller$^d$,
M.~Haguenauer$^d$, 
W.~Hain$^e$,
P.~Hamal$^b$,
K.~Hiller $^{e}$\thanks{Corresponding author, Karl-heinz.hiller@desy.de},
W.~Iwanski$^g$,
S.~Jakobsen$^f$,
C.~Joram$^d$,
U.~K\"otz$^e$,
K.~Korcyl$^g$,
K.~Kreutzfeldt$^h$,
T.~Lohse$^j$,
A.~Maio$^{c1,2}$,
M.J.P.~Maneira$^{c1,3}$,
D.~Notz$^{e,*}$,
L.~Nozka$^b$,
A.~Palma$^{c1,2}$,
D.~Petschull$^e$,
X.~Pons$^d$,
P.~Puzo$^a$, 
S.~Ravat$^d$,
T.~Schneider$^d$,
L.~Seabra$^{c1}$,
T.~Sykora$^i$,
R.~Staszewski$^g$,
H.~Stenzel$^h$,
M.~Trzebinski$^g$,
S.~Valkar$^i$,
M.~Viti$^e$,
V.~Vorobel$^i$,
A.~Wemans$^{c1,3}$\\
\llap{$^a$} LAL, Univ. Paris-Sud, CNRS/IN2P3, Université Paris-Saclay, Orsay, France\\
\llap{$^b$} Palacký University, RCPTM, Olomouc, Czech Republic\\ 
\llap{$^c$} ~$^1$Laboratório de Instrumentação e Física Experimental de Partículas - LIP, Lisboa;
$^2$Faculdade de Ciências, Universidade de Lisboa, Lisboa;
$^3$Dep Fisica and CEFITEC of Faculdade de Ciencias e Tecnologia, Universidade Nova de Lisboa, Caparica, Portugal\\
\llap{$^d$} CERN, Geneva, Switzerland\\
\llap{$^e$} DESY, Hamburg and Zeuthen, Germany\\
\llap{$^f$} Niels Bohr Institute, University of Copenhagen, Kobenhavn, Denmark\\
\llap{$^g$} Institute of Nuclear Physics Polish Academy of Sciences, Krakow, Poland\\
\llap{$^h$} II Physikalisches Institut, Justus-Liebig-Universität Giessen, Giessen, Germany\\
\llap{$^i$} Faculty of Mathematics and Physics, Charles University in Prague, Praha, Czech Republic\\
\llap{$^j$} Department of Physics, Humboldt University, Berlin, Germany\\
\llap{$^*$} Deceased
}

\abstract{The ATLAS Roman Pot system is designed to determine the total
proton-proton cross-section as well as the luminosity at the Large Hadron
Collider (LHC) by measuring elastic proton scattering at very small angles.
The system is made of four Roman Pot stations, located in the LHC tunnel in a 
distance of about 240~m at both sides of the ATLAS interaction point. 
Each station is equipped with tracking detectors, inserted in Roman Pots
which approach the LHC beams vertically. The tracking detectors consist of  
multi-layer scintillating fibre structures readout by Multi-Anode-Photo-Multipliers.}

\keywords{Large detector systems for particle and astroparticle physics;
Particle tracking detectors;
Scintillators and scintillating fibres and light guides;
Performance of High Energy Physics Detectors}

\begin{document}
\include{sections/Introduction}

\include{sections/Detectors}

\include{sections/Electronics}

\include{sections/Testbeam}

\include{sections/Experimental_setup}

\include{sections/DCS}

\include{sections/Trigger_DAQ}

\include{sections/Operation_experience}

\include{sections/Summary}

\acknowledgments
We are indebted to CERN for the 
contributions to the construction and installation of the ALFA detector and its infrastructure.
We acknowledge the help of all technicians and engineers in the
collaborating institutions without whom the ALFA detector could not
have been built. Explicitly we thank for the invaluable contributions during the design, construction,
installation and commissioning phases of the Roman Pot stations:
C.~Cheikali, J.-L.~Socha, D.~Cuisy, C.~Sylvia, B.~Lavigne, M.~Gaspard from LAL Orsay;
C.~David, J.-M.~Lacroix, J.~Noel, B.~Salvant, M.v.~Stenis from CERN; 
P.~Pohl from DESY; T.~K\"onig, M.~Szauter from the JLU Giessen and 
M.~Daniels, M.~Jablonski from the HU Berlin. 

Furthermore we are grateful to all the funding agencies which supported generously 
the construction and the commissioning of the ALFA detector and also provided the 
computing infrastructure.
We acknowledge the supports from FCT Portugal grant CERN/FIS-NUC/0005/2015
and from Polish National Science Centre grant UMO-2015/18/M/ST2/00098.

\end{document}

%% file: sections/Introduction.tex
\section{Introduction}
\label{sec:Intro}

This article describes all aspects of the ALFA detector which is part of the 
ATLAS~\cite{bibIntroduction1} experiment at the LHC.
The ALFA detector was constructed in order to give ATLAS the possibility to detect  
elastically scattered  protons, and to some extent also diffractively scattered protons, 
in the very forward direction. 
Elastic scattering and diffractive physics have an interest in itself. However, elastic 
scattering in the very forward direction can also be used to measure 
the total cross section, $\sigma_{tot}$, and  the luminosity, $L$,  at a collider.
Actually the acronym ALFA stands for 
\textbf{A}bsolute \textbf{L}uminosity \textbf{F}or \textbf{A}TLAS~\cite{bibIntroduction2}.
Traditionally, measurements  of $\sigma_{tot}$ and  the luminosity, using elastic scattering, 
have been done at every new  hadron  collider since the time of the ISR~\cite{reviews}.
 
The  optical theorem connects the total cross section, $\sigma_{tot}$, to the elastic 
scattering amplitude, $f_{el}$, and states:
\begin{equation}
\sigma_{tot}=4\pi \, \mbox{Im} \, [ f_{el}\left(t=0\right)] \,
\label{optical}      
\end{equation}
\noindent where  $f_{el}(t=0)$ is the elastic scattering amplitude in the the forward 
direction  and \textit{t} is  the four-momentum transfer.
The use of the optical theorem requires thus measurements at very small \textit{t}-values 
in order 
to reduce the model-dependent impact from the extrapolation to $t=0$.
Measuring very small \textit{t}-values means in turn measuring the protons at very small 
angles. A special LHC beam optics is needed for this purpose.
The beam has to be more parallel than normally and as the divergence goes as 
$1/\beta^*$special runs with a high $\beta^*$-optics are 
needed.\footnote{The $\beta$-function determines the width of the 
beam envelope around the ring and depends on the focusing properties of the
magnetic lattice. The value at the interaction point is denoted by $\beta^*$.}
To be sensitive to small angles it is necessary to place the detectors far away from 
the interaction point and close to the beam.

Usually Roman Pots~\cite{bibIntroduction3} have been used at all $pp$ or  $ \bar{ p}  p$  
colliders to position detectors very close to the circulating beam. 
A Roman Pot is basically a vessel for detectors that is connected to the accelerator 
vacuum via bellows.
Such an arrangement allows the detector to approach the beam very close without 
entering the machine vacuum.

It is of importance that the detector inside the Roman Pot adds only a minimal amount of 
insensitive material towards the beam to avoid acceptance losses at small $t$-values.   
ATLAS has chosen to equip the Roman Pots with tracking detectors based on scintillating 
fibres. This option has two advantages: the fibres are sensitive up to their physical edge, 
and they cannot interact with induced electro-magnetic noise from the circulating beams. 
 
The ALFA detector was used in LHC Run~1 and, after some upgrades, also in Run~2.
Run~1 was the first data taking period of LHC from 2009 to 2013 followed by Run~2 starting 
in spring 2015.
 
The paper is organized in the following way: 
the construction of fibre detectors, trigger counters, Roman Pots and the photo-detectors
is outlined in section~\ref{sec:Detectors}. The front-end electronics is described in 
section~\ref{sec:Electronics}. In section~\ref{sec:Testbeam} the detector performance 
measured in test beam campaigns is summarised. The experimental set-up in the LHC tunnel 
and the positioning system are discussed in section~\ref{sec:exp}. Section~\ref{sec:DCS} 
deals with the Detector Control System. The triggering and data acquisition systems are 
described in section~\ref{sec:Trigger} while operation experience and upgrades are discussed 
in section~\ref{sec:Operation}. Finally, a summary is given in section~\ref{sec:Summary}.

%% file: sections/Detectors.tex
\section{Detectors}
\label{sec:Detectors}

\subsection{Concept and technology choice}
\label{sec:concept}
The design of the ALFA detector is driven by the goal to precisely measure the track of scattered 
protons at millimetre distance from the LHC beams. 
This is achieved by placing detectors in Roman Pots (RPs), i.e. thin walled vessels which 
allow to operate the detectors inside the LHC beam pipe.
The physics requirements demand a detector resolution of about $30~\mu$m in both coordinates transverse 
to the beam direction with a uniform efficiency up to the detector edge. 
Furthermore, for precise alignment, the distance between the upper and lower detectors needs to be 
measured. 
An Overlap Detector (OD) was built to get this quantity with a precision of 10~$\mu$m.  
\begin{figure} [ht]
  \centering
  \includegraphics[angle=0,scale=0.40]{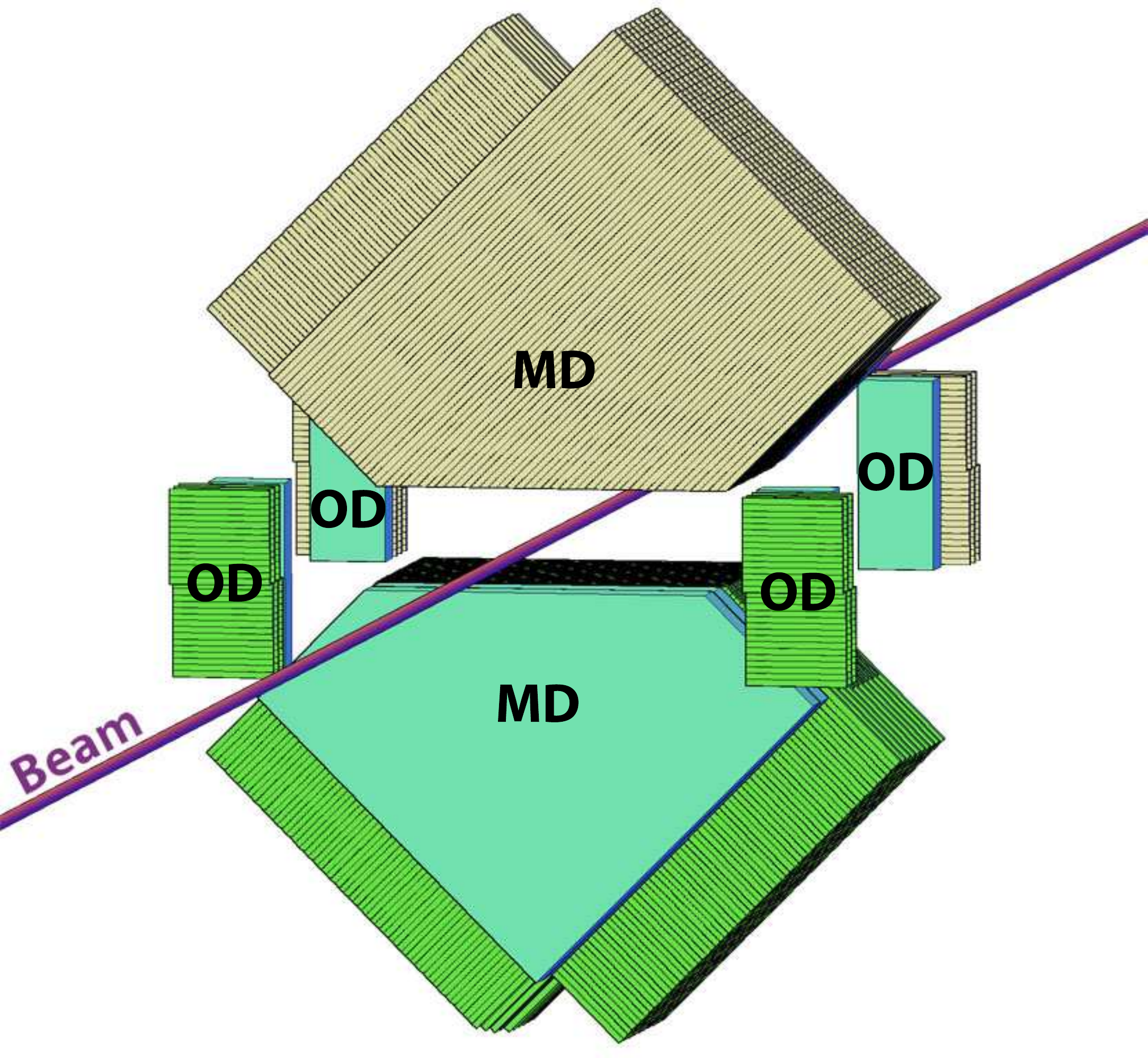}
  \caption[]{A sketch of the detector principle: MD labels the main detectors above and 
   below the beam while OD stands for the overlap detectors. The lines across
   MDs and ODs indicate the orientation of the scintillating fibres. The areas without a fibre structure 
   mark the trigger tiles.
   Copyright CERN: reproduction of this figure is allowed as specified in the CC-BY-4.0 license.}
  \label{fig:Detector-1}
\end{figure}

The ALFA tracking detectors are based on staggered layers of square-shaped scintillating fibres, 
read out 
by Multi-Anode Photo-Multiplier Tubes (MAPMTs). As illustrated in figure~\ref{fig:Detector-1}, 
the fibres of the Main Detectors (MDs) are 
inclined by $\pm 45^\circ$ with respect to the vertical direction. The sensitive region has the shape 
of a truncated square. 
In the ODs the fibres are arranged horizontally measuring the vertical coordinate only.

The scintillating fibre technology has some advantages: the signals are of optical 
nature which provides a maximum immunity to pick-up Radio Frequency (RF) noise from the pulsed LHC beam, 
it does not dissipate any power which facilitates the operation under vacuum,
the detector edge facing the LHC beam is fully active and the staggered multi-layer structure allows to 
tune the resolution to the needs. 
On the other hand, the use of scintillating fibres at a 
hadron collider is a challenge. Organic scintillators provide limited hardness to ionising 
radiation and, particularly for small diameter fibres, the small signal amplitudes 
demand more effort to assure a high and uniform efficiency. 

\subsection{Scintillating fibres}
\label{sec:fibres}
The scintillating fibres, used in the MDs and ODs are of the blue-emitting type.\footnote{Kuraray 
SCSF-78, specification at http://Kuraraypsf.jp/.}
They are square-shaped with outer dimensions of 0.5 $\times$ 0.5~mm$^2$ and have a single 10~$\mu$m 
cladding. The signals have the typical fast decay time of organic scintillators, 2.8~ns. 
The optical attenuation of the signal along the fibre is described by the superposition of two 
exponential functions 
$exp(-x/\Lambda)$ 
with $\Lambda$ = 120~cm and 160~cm, respectively, and plays a secondary role given a fibre length 
of about 35~cm. To minimize cross-talk of primary ultraviolet scintillation light 
between the closely packed fibres, the side faces are coated with a 100~nm reflective 
aluminium film.\footnote{Vacuum coating by physical vapour deposition.}  
In order to maximize the light signals for the photo-detectors, the  $90^\circ$ end faces of the
fibres are mirror coated by sputtered aluminium. 
Due to the machining of the fibre modules, as described in section~\ref{sec:Assembly},
a large fraction of fibres was cut at $45^\circ$ and lost the reflective end face. 
However, more than half of the light arrives to this end face beyond the critical angle of total 
reflection and is sent back to the MAPMTs.

\subsection{Main detectors}
\label{sec:main_detectors}
\begin{figure} [ht]
  \centering
  \includegraphics[angle=0,scale=0.40]{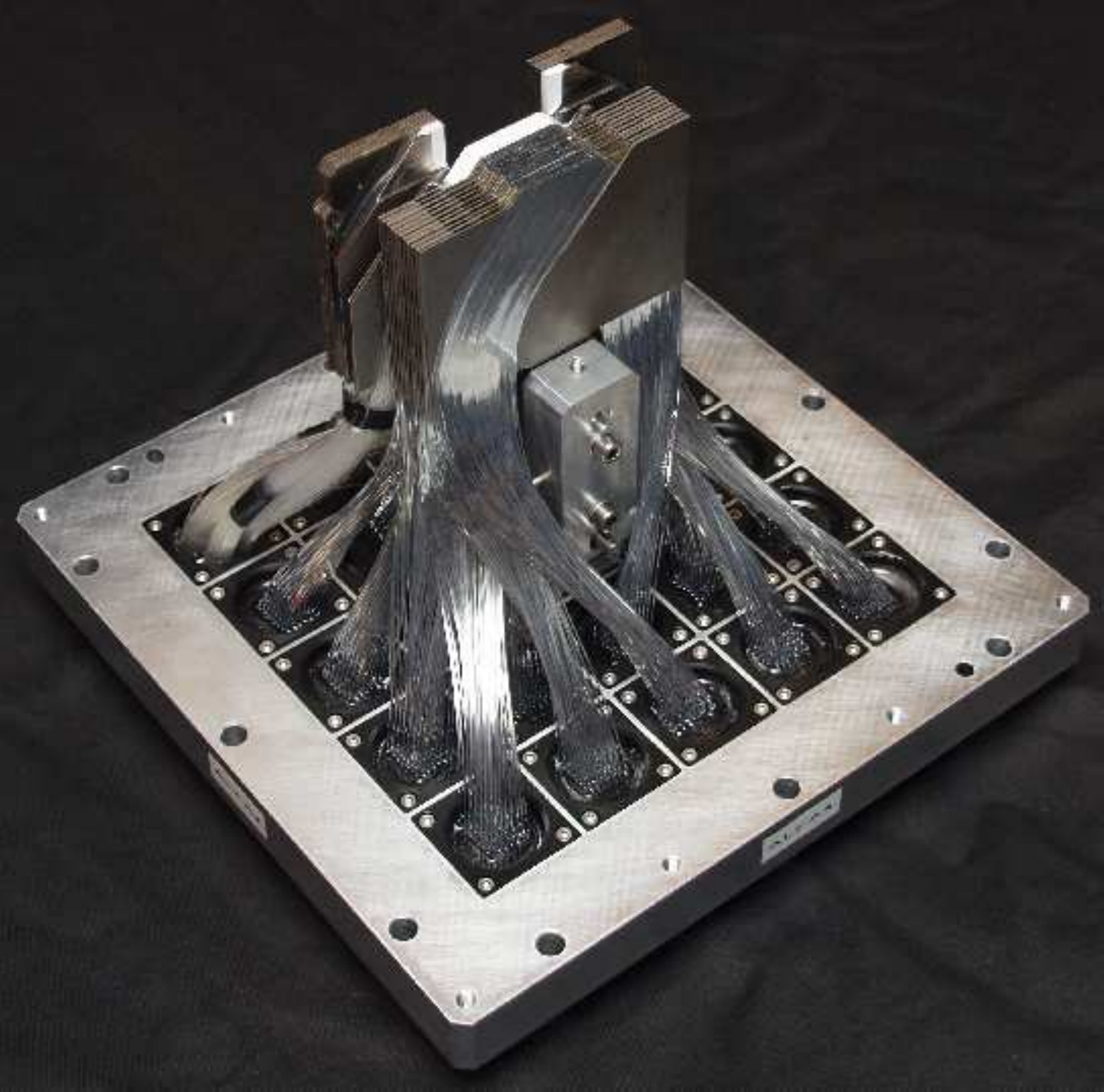}
  \caption[]{An ALFA detector, comprising ten MD modules, three OD modules and the trigger
   counters coated with white reflective paint in between. The scintillating fibres are routed to 
   plastic connectors in the grid of the base plate which on the other side is facing the MAPMTs. 
   The four bundles of clear fibres for the readout of the trigger tiles are routed to the 
   connector in the upper left hole of the base plate.}
  \label{fig:Detector-2}
\end{figure}

Each MD consists of ten fibre modules with two layers of 64 orthogonally arranged fibres. 
These modules exist in ten flavours with layers staggered in steps of a tenth of the fibre size. 
Ignoring the inactive cladding material, the theoretical resolution of a single layer 
of 0.5~mm square fibres is 0.5~mm /$\sqrt{12}$ = 144~$\mu$m. 
Provided all ten layers are precisely staggered and fully efficient, such an arrangement is in 
principle capable to provide a resolution of 14.4~$\mu$m. 
The fibres are tilted by $\pm 45^\circ$ with respect to the vertical coordinate. 
This arrangement avoids that inactive cladding material forms a dead zone at the detector edge. 
Another practical side effect is that 
routing the fibres towards the MAPMTs does not require too small bending radii. 
The details of the assembly procedure are described in section~\ref{sec:Assembly}. 
The various metrology activities for the fibre modules, the detectors and the RPs are documented in 
sections~\ref{sec:survey_fiber} and \ref{sec:pots_surveys}. 
A complete ALFA detector is shown in figure~\ref{fig:Detector-2}. 

\subsection{Overlap detectors}
\label{sec:overlap_detectors}
Each OD consist of three fibre layers with 30 fibres each, which are staggered in steps of a third of the 
fibre size. 
Due to the limited space inside the RPs, the routing of the fibres towards the 
photo-detectors requires a bending by 90$^\circ$ with radii $\le 20$~mm. 
To assure the integrity of the fibres they were pre-bent at a temperature of 85$^\circ$C. 
The ODs are fixed to the same support structure as the MDs, described in section~\ref{sec:Assembly},
They exceed the edge of the MDs in the vertical direction by about 9~mm.
Consequently, the active areas of the ODs begin to overlap when the upper and lower MDs approach 
closer than 18~mm.
The distance can be calculated from a large sample of particles traversing the overlap region of 
upper and lower ODs.
The OD positions with respect to MDs were calibrated in test beam measurements described 
in section~\ref{sec:Testbeam}. 

\subsection{Trigger counters}
\label{sec:trigger_counters}
Both, MDs and ODs, are equipped with scintillator tiles for triggering. 
They are made of 3~mm thick standard plastic scintillator and read out via bundles of clear 
plastic fibres of 0.5~mm diameter.\footnote{Bicron BC-408 or ELJEN EJ-200.}  
The MDs are equipped with two trigger tiles which have the shape of the sensitive detector area.
The ODs have single trigger tiles with a size of 15 $\times$ 6~mm$^2$. 
To maximise the light yield and suppress optical cross talk all trigger
tiles are covered by white reflective paint.\footnote{Bicron BC-600.} 
The signals of the trigger tiles are fed into the trigger system described in 
section~\ref{sec:Trigger}.

\subsection{Detector assembly}  
\label{sec:Assembly}
\begin{figure}[ht!]
  \centering
  \includegraphics[angle=0,scale=0.12]{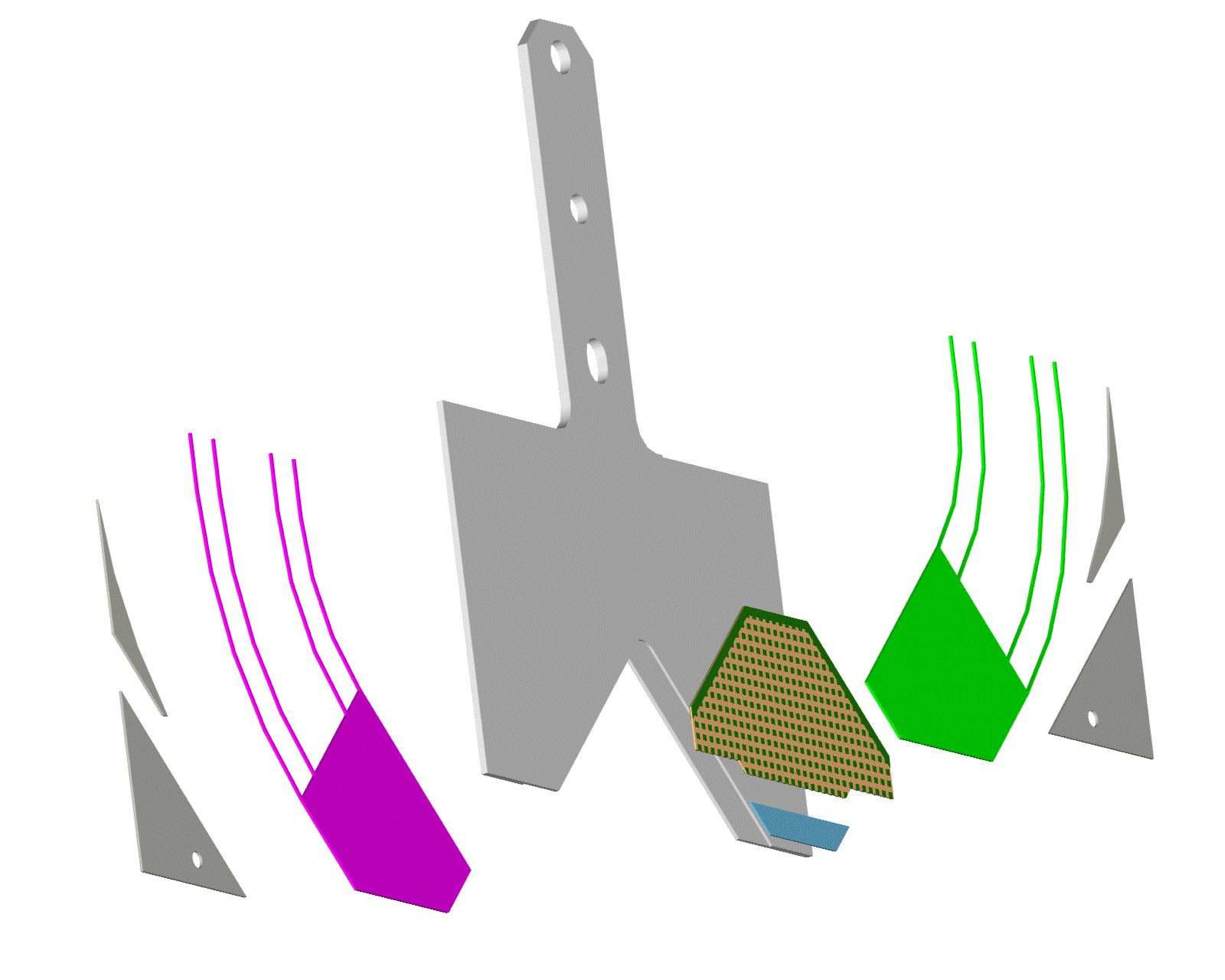}
  \caption[]{Exploded view of the module assembly procedure.
   The central part (grey) is the titanium substrate with two precision holes for positioning
   and another hole for the fixation screw.
   From the inner to outer positions follow the G10 sheet (yellow with green dots), 
   the acryl sheet (blue), two stacks of 64 scintillating fibres (pink, green) and the wedges 
   (grey) to support the fibre routing.}
  \label{fig:assembly}
\end{figure}

The assembly of the MD fibre modules proceeded in several steps as illustrated in 
figure~\ref{fig:assembly}. 
A titanium plate with precisely machined 45$^\circ$ angles for the fibre positioning
served as a base substrate. To facilitate the gluing of the fibres a G10 sheet with lithographic 
pattern was glued between the two angles of the titanium plate filling the empty space of 
the fibre crossing area.\footnote{G10 is a fibre glass epoxy composite laminate.}    
On each side of the substrate 64 fibres were sequentially positioned starting from the  
45$^\circ$ precision angles.
In their final positions the fibres were fixed by several auxiliary wedges and clamps,
the module was turned and the other side assembled by the same procedure. 
Then the module was brought in an upright position and the gluing process started by 
immersion of the fibre ends in a tank of glue.
The glue dispersed by capillary effect in the small gaps between the substrate and the fibre stack.
Before completion of polymerization the fibre ends were removed from the tank. The excess of glue 
was removed and in the area where the fibres are not supported by the 
G10 plate an acryl sheet was inserted to ease the subsequent machining of this edge. 
In the final step of the gluing procedure the fibres in the crossing area were kept straight by
corresponding wedges shown in figure~\ref{fig:assembly}. Outside of the crossing area the 
fibres were bent to meet the connectors towards the MAPMTs.
The 30 fibres of an OD module were glued in batches of 15 fibres to both sides of the 
titanium substrate. The glue was directly dispersed on the substrate and the fibres
were arranged one-by-one on the prepared area fixed with auxiliary wedges.  

The MD and OD trigger counters were produced by gluing the scintillator tiles on substrates with 
0.5~mm lips while ensuring the position of the lower edge by a precision jig.
Clear lightguide fibres were assembled to bundles and glued to the scintillator tiles.

After the module assembling the fibre positions were measured by microscope,
described in section~\ref{sec:survey_fiber}. Then the modules were machined at the  
lower edge to get a distance of 135~mm to the precision hole of the 
titanium substrate indicated in figure~\ref{fig:Survey-md-metrology}.

\begin{figure} [ht]
  \centering
  \includegraphics[width=50mm]{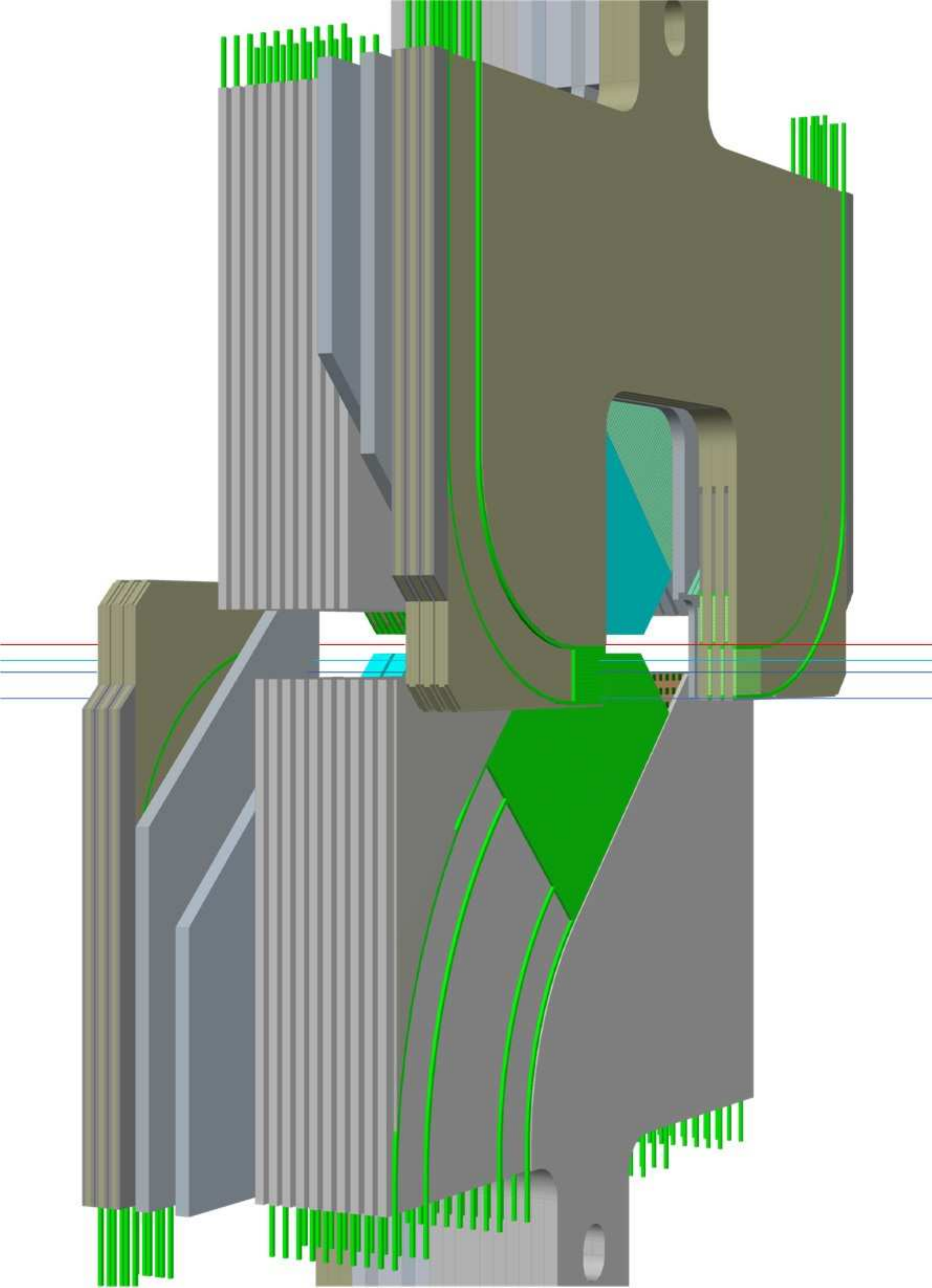}
  \caption[]{A schematic sketch showing the composition of fibre and trigger modules in an upper 
  and lower RP. 
  From outside to inside: three OD modules (dark grey), the OD and MD trigger counters (blue) and 
  ten MD modules (light grey). 
  The sensitive areas of the upper and lower ODs overlap at both sides of the MDs.  
  The small gap between the upper and lower MDs ensures the passage of the LHC beam.}       
  \label{fig:detector_full}
\end{figure}

Next, ten MD modules, three OD modules and three trigger modules were assembled to the full
detector composition.
They were stacked on two precision pins of the central support arm which is attached to the 
detector base plate.
The support arm has adjustable elements to optimize the clearance of the detector to the 
RP walls and to optimize the distance to the inner RP bottom window. 
It also incorporates a cable feedthrough for temperature probes and 
Light-Emitting Diodes (LEDs). 
A schematic 3D view of the module structure is shown in figure~\ref{fig:detector_full}.

Finally, the scintillating fibres were
routed to the feedthroughs in the base plate and inserted in the connector holes.
The connector bowl was filled with glue and after polymerization the side facing 
the MAPMT was machined with a diamond tool and polished to ensure a good optical contact.

\subsection{Fibre position survey}
\label{sec:survey_fiber}
The resolution of multi-layer fibre detectors depends on the staggering of the individual
layers. The staggering value of each layer is determined by the position of the 
first fibre attached to precision groove of the titanium substrate. Due to effects of machining
precision and deviations of fibre widths from the design value, the fibre positions 
differ from the nominal values. The impact on the detector resolution can be 
partly compensated if the real fibre positions are used for the track reconstruction.   
For this purpose all fibre positions were measured by a 
2D microscope.\footnote{Nikon DS-5M Digital Camera.}
\begin{figure} [ht]
  \centering
  \includegraphics[angle=0,scale=0.70]{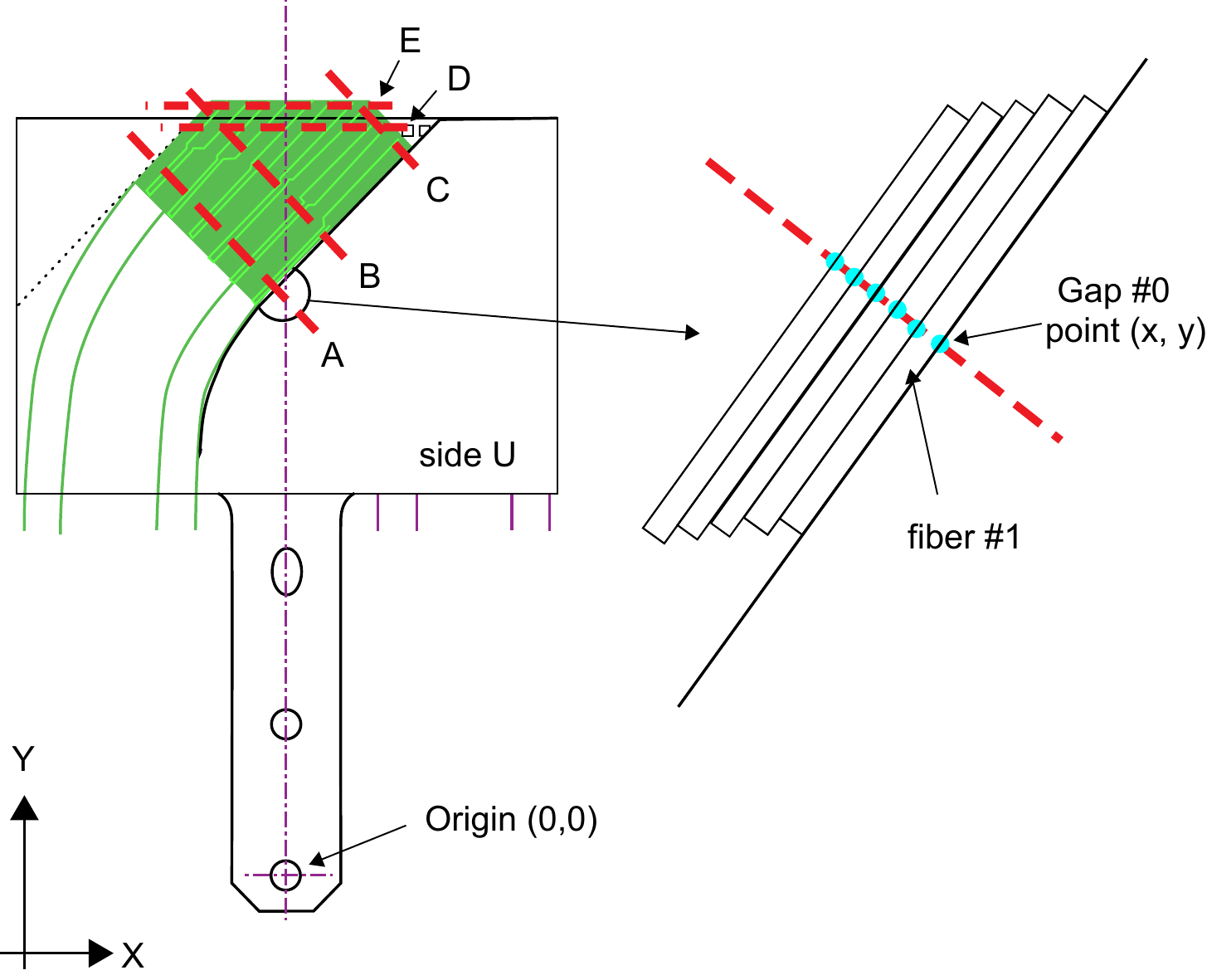}
  \caption[]{\small Sketch of a fibre substrate with the lines A to E where the 
   gaps between the fibres 
   are measured. The detector coordinate system is defined by the precision hole (0,~0)
   and the long hole towards the fiducial detector area.}
  \label{fig:Survey-md-metrology}
\end{figure}
The measurement scheme for the MD fibre substrates is shown at figure~\ref{fig:Survey-md-metrology}. 
From the gap positions at both fibre edges along the lines A to E, the central values were calculated 
and fitted to straight lines. 
The deviations from the nominal fibre positions depend on the fibre number and differ  
along the measured lines.
They can accumulate up to a few 100~$\mu$m and are specific for each substrate.
The measurement scheme for the OD substrates is based on the same principle. 

\subsection{Roman Pots and related surveys}
\label{sec:pots_surveys}
\begin{figure}[ht]
  \centering
  \includegraphics[angle=0,scale=0.60]{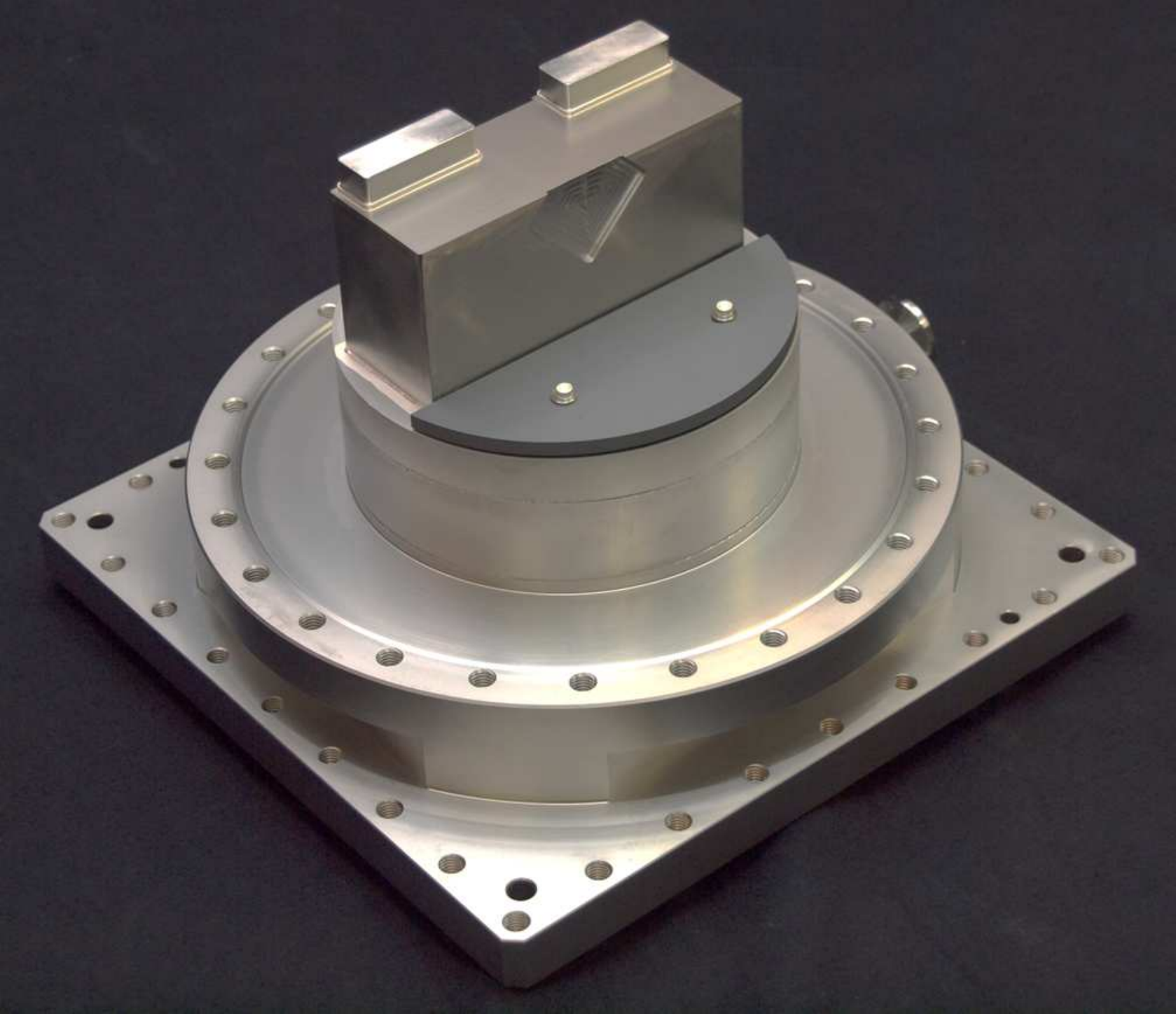}
  \caption{A photo of a bare Roman Pot.
    In uppermost positions extending the bottom window are the two extrusions for the ODs.
    At the side wall the front window machined to the sensitive MD area is visible.
    The ferrite tiles are attached to the cylindrical RP bottom. The square flange which is 
    foreseen to hold the detector has outer dimensions of 264 $\times$ 264~mm$^2$.}        
  \label{fig:RP_3D}
\end{figure}

The RP houses the detector and  therefore its design takes into account the constraints 
imposed by the detector geometry as well as the compatibility with the movement system of the station.
A RP is shown in figure~\ref{fig:RP_3D}. It consists basically of a cylindrical pot 
with special caverns for one MD and both ODs and two flanges. The circular flange is connected to the 
bellow which is attached to the LHC vacuum chamber 
and allows the movement of the RP.
The squared flange is used to fix the detector base plate with the feedthroughs to the 
MAPMTs and trigger Photo-Multiplier-Tubes (PMTs). 
The RP is mostly made of 2~mm stainless steel. The side walls perpendicular to the beam
were machined to 0.5~mm thickness and the RP bottom window was reduced to 200~$\mu$m.
At the bottom of the cylindrical part of the RP a semi-circular ferrite tile is attached. 
It aims to absorb the electro-magnetic power of the RF losses related to the impedance of 
RP cavities. 
Although the fibres themselves are insensitive to electro-magnetic radiation, the RF losses result in 
a heating of the titanium plates where the fibres are glued. This heating effect and preventing
measures are described in section~\ref{sec:RF}.

Due to the thin 200~$\mu$m RP bottom window the protection of the ultra-high LHC vacuum
is a major safety aspect. For this purpose and to minimize the deformation induced by
the atmospheric pressure on the window the interior of the RP is kept in a secondary vacuum.    
Pumps close to the stations keep the pressure below 20~mbar.
Pressure tests have shown that the thin RP bottom window withstands overpressure up to 
80~bar before destruction. Since the maximum encountered overpressure in case of an accidental 
loss of the secondary vacuum would be 1 bar the ultra-high LHC vacuum is well protected.   
To maintain the secondary vacuum all fibre feedthroughs of the base plate which closes the RP 
need to be carefully  sealed by glue in the last step of the assembling procedure described in 
section~\ref{sec:Assembly}.
For the precise positioning and to ensure the necessary clearance all RPs and detectors 
were surveyed with a mechanical 3D measuring device
with a precision of about 5~$\mu$m.\footnote{Olivetti Inspector MAXI 900V.}
The distance to the RP bottom window was measured at five positions close to the corners and 
in the centre. A small outward bend of about 50~$\mu$m was observed at the central position.
 
Before the detectors were assembled, all fibre modules were measured and grouped 
in batches with similar module length.
After assembling a final 3D measurement of the lower detector edge was performed. 

\begin{figure} [ht]
  \centering
  \includegraphics[width=150mm]{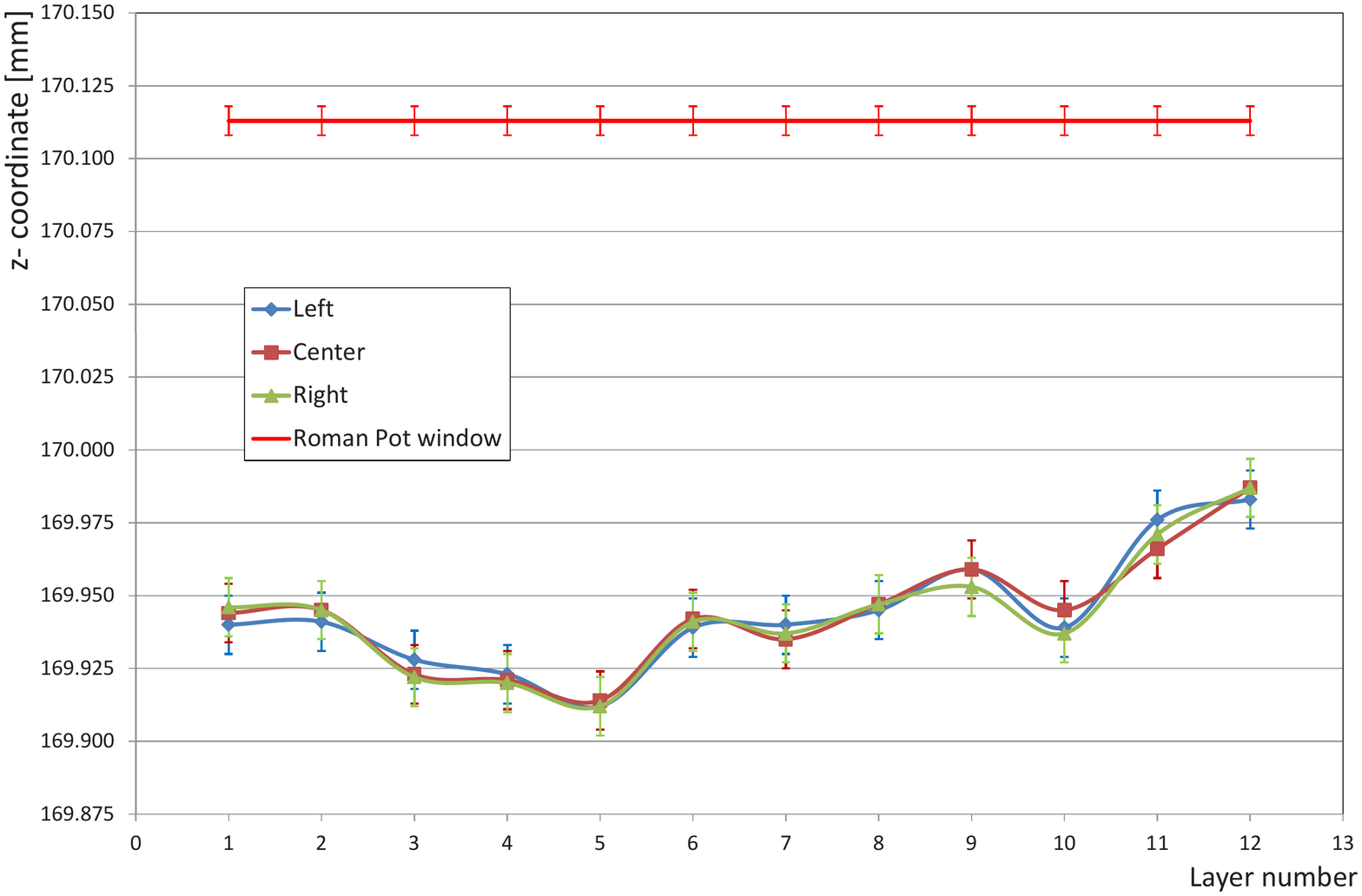}
  \caption[]{The positions of the lower edges of the MD modules (1-10), the MD trigger 
   tiles (11-12), and the inner RP window (red line) are indicated with respect to the RP flange. 
   For all modules three values are quoted: in left, central and right positions along
   the lower detector edge.}
  \label{fig:Survey-fiber+trigger_3D}
\end{figure}

For a typical detector the results are shown in figure~\ref{fig:Survey-fiber+trigger_3D}. 
The trigger tiles, which are covered by white reflective 
paint of thickness between 20~$\mu$m and 40~$\mu$m, 
are slightly protruding.
To ensure an optimal gap between the lower detector edge and the 
inner RP bottom window, the detector fitting best in each individual RP was selected.
All gaps range between 150~$\mu$m and 250~$\mu$m. 
Under vacuum conditions, the expected bend of the base plate which carries the detector
is in the order of 40~$\mu$m, reducing the gap by this amount. 
Before the insertion of the detectors in the RPs the clearance to the inner RP walls 
was adjusted by small screws impacting the angles of the 
detectors on the holding arm. The typical clearance 
was between 0.5~mm and 1~mm, in a few extreme cases reduced to a few 100~$\mu$m. 

\subsection{Photo-detectors}
\label{sec:photodetectors}
The MAPMT technology, a reliable and proven technology, 
fulfills all requirements for the readout of a relatively large number of scintillating fibres. 
The chosen device
provides an $8 \times 8$  grid of channels with a nominal 
pitch of 2.3~mm, including 0.3~mm gaps for separations of the sensitive readout pixels, which 
is adequate for the readout of 0.5~mm square fibres.\footnote{Hamamatsu R7600-00-M64, specification 
http://www.hamamatsu.com/.}
Even though the cross-talk between direct 
channels is small, the fibres were routed such that neighbouring fibres were not read by 
neighbouring MAPMT channels. The typical quantum efficiency is 26\% and the gain is 
in the order of $10^6$ at a bias of -1000~V. However imperfections of the micro-machined dynode 
channels lead to channel-to-channel variations of a factor 2-3. The average gains of different 
MAPMTs can differ by a similar factor. 

For technical reasons, the internal dynode structure is displaced by about 1~mm relative to 
the tube centre. The displacement can be obtained from alignment marks on the first dynode and 
corrected for by gluing plastic shims on the MAPMT bodies. 
Due to its compact size, this MAPMT operates in the earth magnetic field without any shielding. 
In view of possible stray fields in the LHC tunnel, all MAPMTs as well as the trigger PMTs 
are protected by a 
mu-metal grid which is integrated in the base plate structure, described in 
section~\ref{sec:Assembly}.\footnote{A nickel-iron soft magnetic alloy with very high permeability 
suitable for shielding against magnetic fields.}

The scintillator tiles of the trigger counters are read out by 8~mm diameter miniature PMTs 
of the metal package type. For the MD trigger tiles a PMT with super-bialkali 
photo-cathode is used which offers a peak quantum efficiency of 35~\%.\footnote{Hamamatsu R9880U-110, 
specification http://www.hamamatsu.com/.} 
The OD trigger tiles, where the detection efficiency is less critical, are read out 
by a PMT with conventional bialkali 
photo-cathode.\footnote{Hamamatsu R7400P, specification http://www.hamamatsu.com/.}

%% file: sections/Electronics.tex
\section{Front-end electronics}
\label{sec:Electronics}

The front-end electronics is fitted in a tight volume on top of the ALFA stations. 
The timing, control and data communications are sent over more than 240~m 
to the Detector Control System (DCS) and Data AcQuisition system (DAQ) 
with adapted techniques 
to ensure the data rate and the latency of the first level trigger (L1) decision.    
The high voltage biasing lines are deployed along the LHC tunnel and through local patch panels
at the stations connected to the MAPMTs or trigger PMTs.  
Each electronics assembly receives two low voltage power lines (5~V, 10~A and 7~V, 1~A) 
delivered from local power units, placed close to the stations  
in a service alcove to avoid excessive voltage drops through the cables. 

\begin{figure}[ht]
  \centering
  \includegraphics[angle=0,scale=1.5]{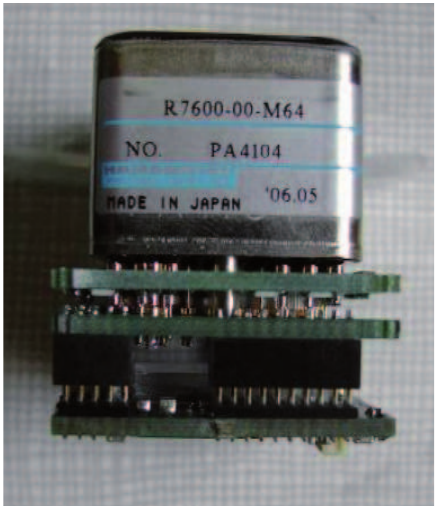}
  \caption[]{A side view of a MAPMT with a stack of three boards, the PMF, 
   which performs the readout of the fibre signals.}
  \label{fig:PMF}
\end{figure}

The Photo-Multiplier Front-end electronics (PMF) is a compact stack of three boards 
connected directly to the backplane of each MAPMT. 
The first board provides the high voltage distribution to the dynodes of the MAPMT.
A transition board is equipped with contact sockets for the MAPMT pins and passes the signals to 
the third board that holds the analogue Multi-Anode ReadOut Chip (MAROC) and a digital readout 
Field-Programmable Gate Array (FPGA).
A PMF attached to the MAPMT backplane is shown in figure~\ref{fig:PMF}.

In the 64-channel MAROC chip~\cite{bibElectronics1}, the MAPMT signals are 
pre-amplified, shaped with 15~ns shaping time and fed into a programmable discriminator.
The global threshold and gain setting per channel are configured 
through the local FPGA from the ALFA DCS. 
The MAROC provides also a 110~ns slow shaper followed by a sample and hold and a multiplexer 
circuit that can be activated to read the analogue charge per channel. 
Due to the multiplexing time, the analogue charge readout is operational at a low rate of a few kHz
and was used only for tests.
\begin{figure} [ht]
  \centering
  \includegraphics[angle=0,scale=0.55]{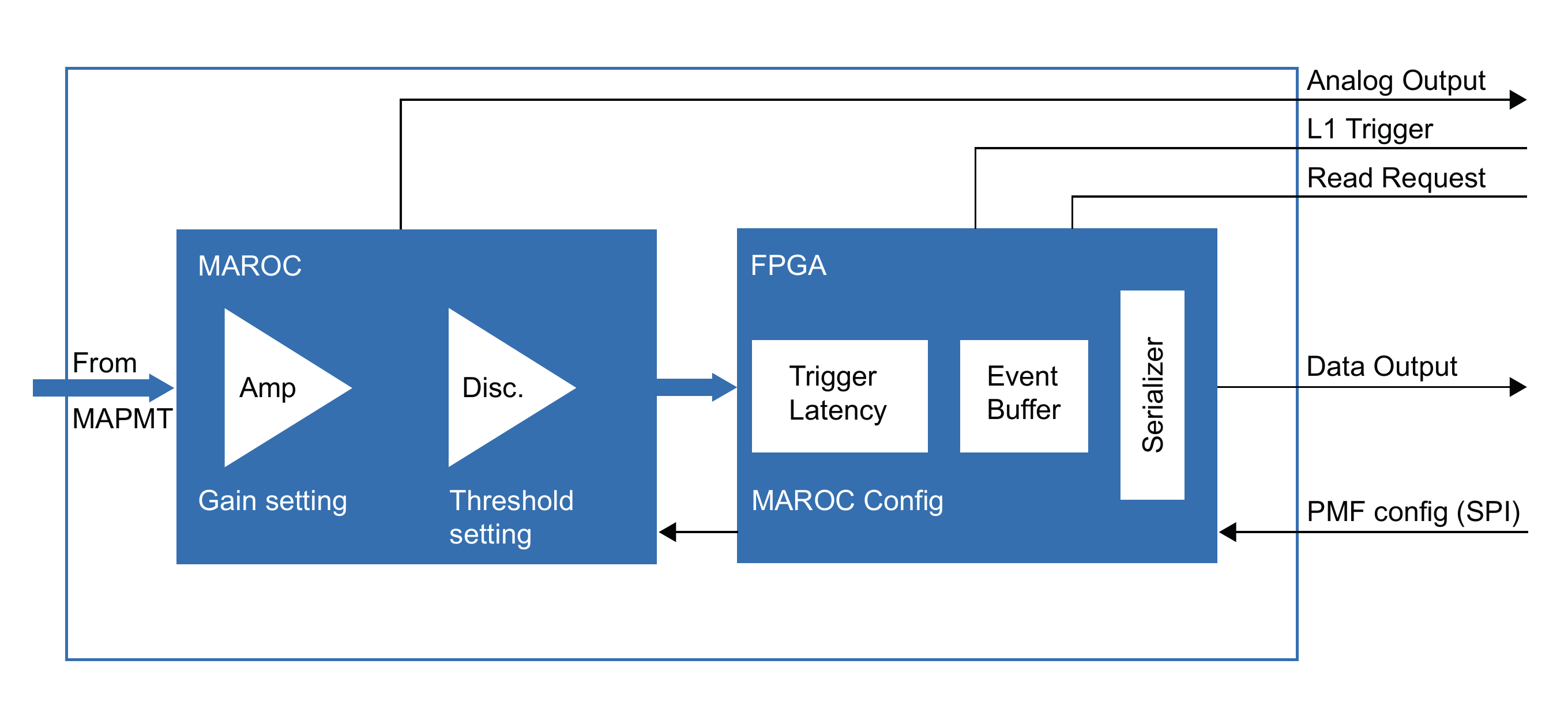}
  \caption[]{The architecture of the front-end electronics for the readout of the fibre signals 
   from the MAPMTs, as it is realised in the PMF.} 
  \label{fig:el_architecture}
\end{figure}
The MAROC 64 discriminator outputs are fed into the FPGA where they are continuously 
sampled at 40~MHz, the LHC bunch crossing rate. 
The sampled data are buffered in a pipeline until the centrally processed L1 trigger 
signal arrives. With the L1 decision the digital data of
the associated bunch crossing are transmitted through a 40~MHz serial interface to another 
FPGA  located on the motherboard, which collects the data from all 23 PMFs per detector.
The general layout of the fibre front-end electronics is shown in figure~\ref{fig:el_architecture}.

\subsection{Motherboard functions}
\label{sec:motherboard}
The motherboard receives the clock, trigger and command signals from the central ATLAS 
Timing, Trigger and Control (TTC) system 
through an optical link. It controls all communication with the 23 PMFs per detector. 
After the L1 decision the PMF data are formatted 
to the event structure defined for the MROD modules of the ALFA branch of the DAQ system, 
described in section~\ref{sec:DAQ}. 
The data are transmitted at 
1.2~Gb/s by a fast serialiser, a gigabit optical link~\cite{bibElectronics3}, and an optical driver. 
The motherboard functions, and all the front-end electronics elements, are controlled and 
monitored by an Embedded Local Monitoring Board (ELMB) interface card~\cite{bibElectronics4} 
which is connected to the ALFA branch of the DCS.

\subsection{Triggerboard functions}
\label{sec:trigger_board}
The triggerboard is mounted as a mezzanine on top of the motherboard.
The four signals of the trigger PMTs, two from the MD scintillator tiles and two from the OD scintillator 
tiles, are pre-amplified, shaped and 
discriminated by two MAROC chips. The binary output of the MAROCs is fed into a FPGA  
that is configured to perform various combination of the four trigger signals. 
The output signal is sent to the counting room to contribute to the L1 decision in the
Central Trigger Processor (CTP), as outlined in section~\ref{sec:Trigger}.

The triggerboard supports two programmable pulse generators, designed to deliver variable amplitudes 
(up to 4~V) and short pulses (5 to 50~ns range) to the two LEDs installed inside each RP. 
The LEDs are placed close to the scintillating fibres and used to generate test signals  
to verify the readout system.

%% file: sections/Testbeam.tex
\section{Detector performance}
\label{sec:Testbeam}

A series of test beam measurements were carried out to assess the performance of the ALFA 
detectors.
In 2005 the principle of a scintillating fibre tracker was demonstrated~\cite{TB_2005}.
In 2010 all but one detector were inserted in two RP
stations equipped with the final version of the front-end electronics. To optimize
the performance data sets with different detector parameters were analysed.

The 2010 beam test was performed in the H6B area of the H6 beam line of the CERN SPS North 
Area with a 120~GeV hadron beam. For precise tracking the EUDET silicon pixel 
telescope~\cite{EUDET} was used. 
It consists of six sensors layers with an internal space point resolution of 4.5~$\mu$m
covering an active area of 21.2 $\times$ 10.6~mm$^2$.
The ALFA stations with the fibre detectors were placed about 1~m downstream of the EUDET 
telescope. At this distance the pointing precision for tracks reconstructed in five or more 
pixel layers ranged from 3~$\mu$m to 5~$\mu$m.

\begin{figure}[h!]
  \centering
  \includegraphics[width=120mm]{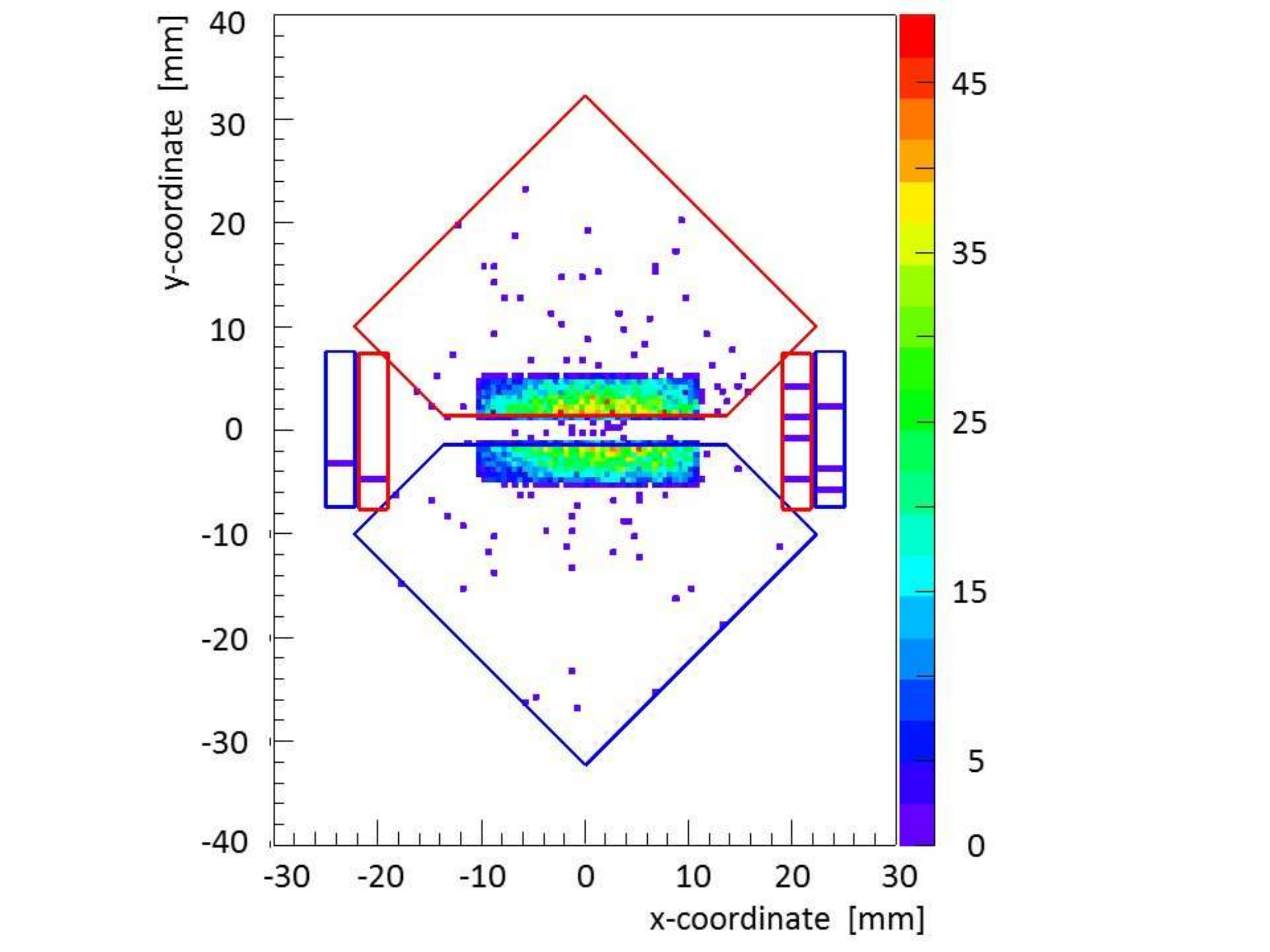}
  \caption{A track pattern with the beam in central position on the station. The distance 
   between upper and lower RPs was about 2~mm. The size of the track pattern 
   is defined by the lateral size of the EUDET trigger counters. At both sides of the MDs the 
   ODs are symbolised.} 
  \label{fig:TB_2010_scan}
\end{figure}

Since the transverse area of the EUDET sensor was smaller than the area of the fibre detectors, 
a position scan in vertical and horizontal directions was performed. 
A typical track pattern where the beam hits the edges of the upper and lower MDs is shown in 
figure~\ref{fig:TB_2010_scan}.  
The tracks reconstructed by the EUDET
telescope were used to measure the distance between the detector edges. 

In the next sections the following performance parameters are documented: 
the light yield of the scintillating fibres 
and of the trigger scintillators, the hit multiplicities and related layer efficiency, the cross 
talk, the spatial resolution, the edge sensitivity and the position calibration of the ODs.
The quoted values are based on optimised detector settings for high-voltages, gains and thresholds 
as they are used later for LHC data taking. 

\subsection{Light yield}
\label{sec:Light_yield}
A high light yield is a key parameter to ensure a good tracking quality. 
The first measurements based on a module of 
2 $\times$ 32 fibres were performed in a DESY test beam with 6~GeV/c electrons~\cite{TB_2005}. 
Later, the fibres of all detectors were exposed to cosmic particles to calibrate 
the amplifiers and equalize the signal amplitudes for discrimination~\cite{Sune_2010}. 

Before the cosmic exposure each fibre channel is illuminated at low light level from a pulsed
LED to determine the mean amplitude $x_1$ and the width $\sigma_1$ of the 
1~Photo-Electron (PE) signal. 
With the known 1~PE parameters the amplitude spectrum $Q(x)$ of cosmic particles can be 
described by a Poisson distribution $P(i,\mu)$ convoluted with the Gaussian functions of all 
$i$~PE signals as described by formula~(\ref{equ:qfunc}):
\begin{equation}
  Q(x) = A \sum_{i=0}^n {P(i,\mu)} \cdot e^{-(x-x_i)^2/2{\sigma_i}^2}
  \label{equ:qfunc}
\end{equation} 
The mean amplitude and the width of the $i$-th PE term are parametrised by the 1 PE signal:
$x_i=i\cdot x_1$ and ${\sigma_i}^2={\sigma_0}^2+i\cdot{\sigma_1}^2$, respectively.
The sum in (\ref{equ:qfunc}) is cut off at $n=15$ PEs.
The fit provides the overall normalization factor $A$, the mean amplitude and the width 
of the pedestal, $x_0$ and $\sigma_0$, and the average number of $\mu$~PEs. 

A striking difference to the beam data is the angular distribution of incident cosmic particles. 
This results in two effects which need to be corrected for an estimate of the light yield of 
beam particles traversing the detectors under 90$^\circ$:
a cosmic particle can deposit more energy per fibre due to the inclined trajectory, 
and for the same reason it can cross more than one fibre.  
Simulations have shown that the first effect increases the light yield by
a factor 1.2, while the second effect enlarges the cross talk by about 20~\%.

\begin{figure}[h!]
  \centering
  \includegraphics[width=130mm]{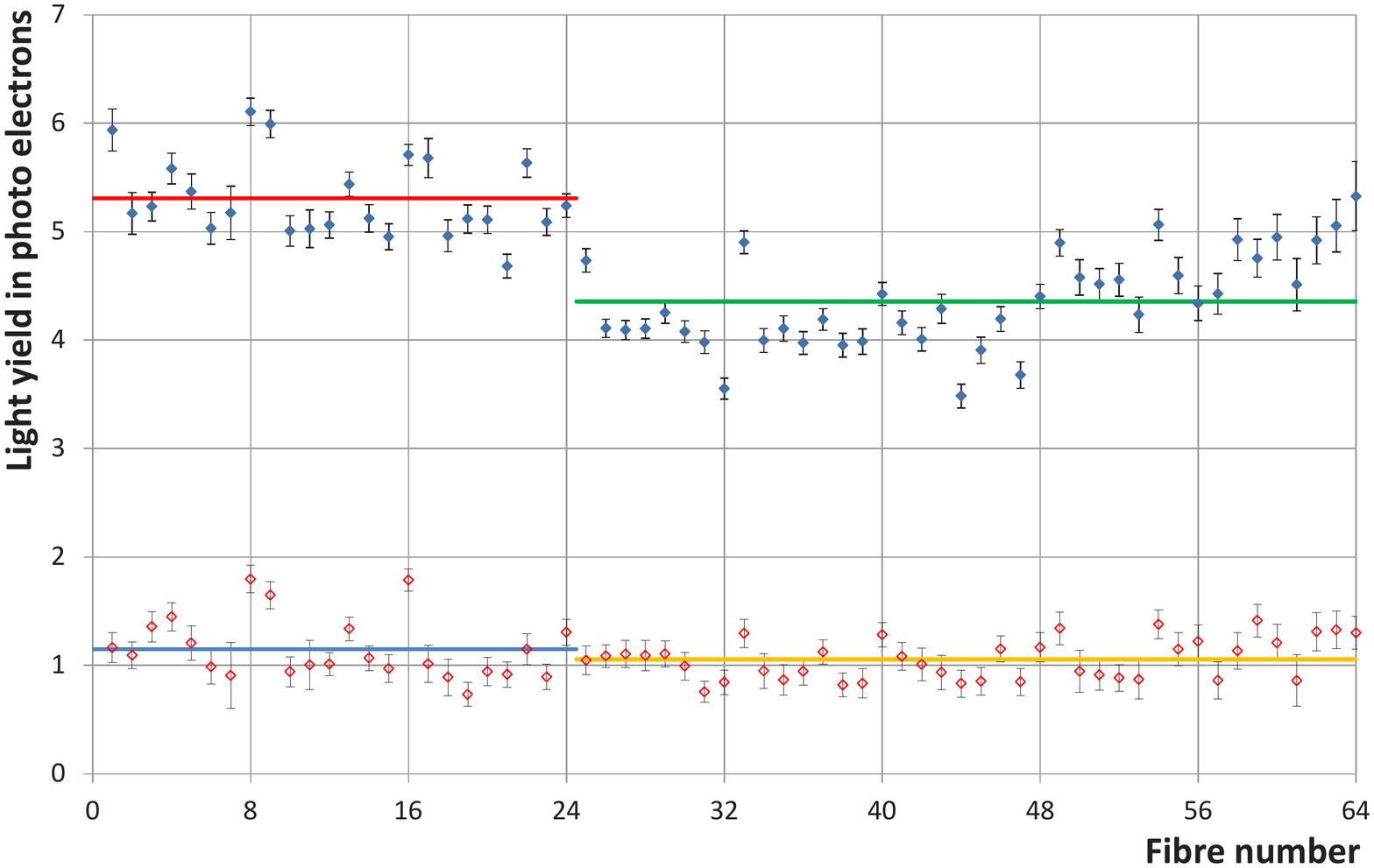}
  \caption{The light yield of a layer of 64 fibres in terms of PEs. The fibres 1-24 
   have a reflective aluminium end face, while the fibres 25-64 were cut and lost the 
   reflective end. The lower data points describe the cross talk contribution due to 
   inclined cosmic particles traversing neighbouring fibres.}   
  \label{fig:light_yield}
\end{figure}

The light yield of a typical fibre layer exposed to cosmic particles is shown in 
figure~\ref{fig:light_yield}.
For the fibres with the aluminium coated end faces the yield is about 5~PEs, while for fibres  
which lost the coating by machining it is 0.5~PE to 1~PE less. 
The average light yield of all layers varies by $\pm$1~PE.
The cross talk generated by inclined trajectories is at the level of 1~PE. 
For the light yield of the OD fibres similar results were obtained. 
According to Poisson statistics a light yield of 4~PEs allows a single fibre efficiency
of 98~\%. 

For the trigger tiles different coatings and light guide version were investigated 
in a DESY test beam with 6~GeV electrons~\cite{Franz_2009}.
The best results were achieved  with tiles covered by white reflective paint read out 
by flexible bundles of round light guide fibres of 0.5~mm diameter. 
The measured light yield was about 40~PEs. 

\subsection{Fibre layer efficiency}
\label{sec:Efficiency}
\begin{figure}[h!]
  \centering
  \includegraphics[width=150mm]{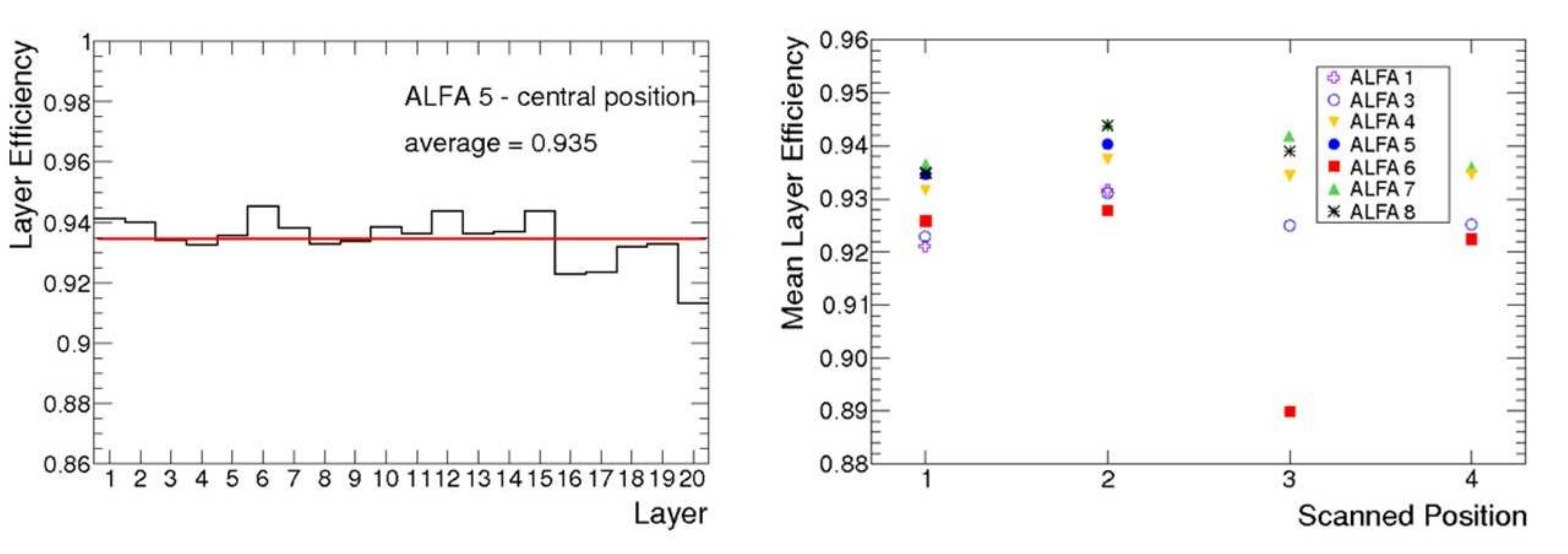}  
  \caption{Left: the layer efficiency of the ALFA5 detector with the beam focused 
   to the central detector area. Right: the layer efficiency of the ALFA1--8 detectors scanned 
   at four positions - the central detector area~(1), the lower detector edge~(2), 
   the left~(3) and the right~(4) detector corners. Apart from an outlier (ALFA6 at position 3)
   all values are centred at the average with a maximum spread of $\pm~1\%$.} 
  \label{fig:TB2010_layeff}
\end{figure}

For each charged particle passing a fibre layer one would expect exactly one fibre hit. 
Due to secondary interactions in the surrounding detector material, electronic noise 
and cross talk the average multiplicity per layer is larger. 
Dependent on the position where the beam hits the detector the values 
range between 1.05 and 1.27 hits per layer. 
To measure the layer efficiency the trajectories of beam particles were reconstructed
from hits in all layers excluding the layer under inspection. 
For a typical MD the layer efficiency with the beam incident to the central 
detector area is shown in figure~\ref{fig:TB2010_layeff}. The average value of 93.5~\% degrades 
slightly along the particle path through the detector due to losses by secondary interactions.
Also shown in figure~\ref{fig:TB2010_layeff}, the layer efficiency varies as a function of   
the incident beam position from 92~\% to 94~\%, with a typical spread of 
about 1~\% among the detectors. 
These values are in good agreement with the expectations taking into 
account efficiency losses due to the 10~$\mu$m insensitive cladding material (-4~\%) and the 
1~PE amplitude threshold (-2~\%) for the digitisation.

\subsection{Cross talk}
\label{sec:cross_talk}
Cross talk in adjacent fibres or MAPMT channels fakes signals which degrade the performance of
a tracking detector. 
The first measurement of cross talk measurements was based on the early DESY data~\cite{TB_2005}. 
Later, measurements with cosmic particles and SPS test beams were done with final
detectors and front-end electronics.
The mapping of fibres to MAPMT channels was optimised to reduce the impact 
of cross talk on the track reconstruction. 
The evaluation of cross talk is based on hit counting, while the early DESY data were analysed 
using amplitude information. Due to these differences, the later SPS test 
beam results are a more realistic estimate of the performance under LHC conditions. 

\begin{figure}[h!]
  \centering
  \includegraphics[width=120mm]{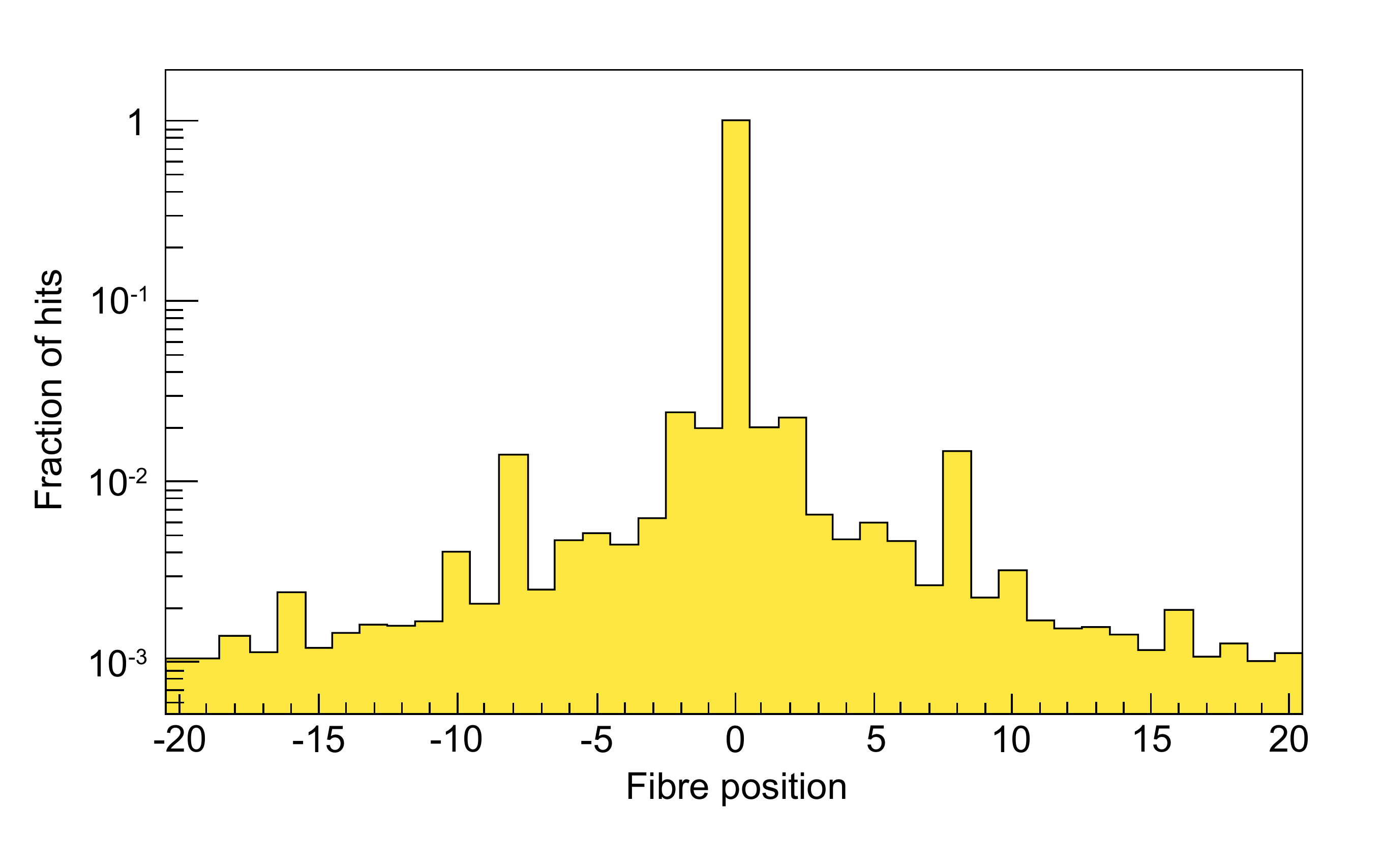}
  \caption{Distribution of fibre hits with respect to the position of the particle trajectory.
    The structure with peaks at various positions, e.g. at $\pm$8 fibres, results from the 
    MAPMT cross talk in combination with fibre-to-MAPMT mapping scheme.} 
  \label{fig:TB2010_cross_talk}
\end{figure}

The relative amount of hits in fibres aside the reconstructed particle track
is  shown in figure~\ref{fig:TB2010_cross_talk}. 
It results from the superposition of four types of cross talk:
the direct fibre cross talk, the MAPMT cross talk, and the cross talks attributed to 
neighbouring connector and MAROC channels.  
The fibre cross talk amounts to an average value of 6~\% and is a mixture of two components: 
2~\% for the fibres with coated end faces and 8~\% for the fibres without coating due to the 
45$^\circ$ machine cut.
The latter enhanced value is attributed to light exiting the fibres towards the RP bottom window
in about 200~$\mu$m distance where it is reflected into the neighbouring fibres with uncoated end 
faces.
To reduce these reflections the inner RP windows were subsequently covered by an anti-reflective 
paint which reduces the amount of reflected light by factors between three and 
five.\footnote{Deflocculant Acheson Graphite.}

The cross talk in neighbouring MAPMT channels is about 4.5~\% for the direct neighbours and 1~\% 
for the diagonal neighbours. According to the fibres-to-MAPMT mapping scheme, 
the cross talk in direct MAPMT channels enhances the amount of hits in a distance of two, eight 
or ten fibres with respect to the genuine track hit, visible in figure~\ref{fig:TB2010_cross_talk}.  
The cross talk related to the MAROC chip and connectors contributes at the level of 2~\% each.

The various cross talks are the dominant source of noise hits which contribute to the hit 
multiplicity per layer, described in section~\ref{sec:Efficiency}.

\subsection{Spatial resolution}
\label{sec:Resolution}
The spatial resolution was measured by extrapolating the tracks reconstructed by the EUDET 
telescope to the position of the MDs. The uncertainty of this extrapolation  
was about 5~$\mu$m if at least five pixel layers were used. 
The track reconstruction in the MDs is based on the overlapping of fibres which 
are attributed to the particle trajectory. 
For this purpose the measured fibre positions, described in section~\ref{sec:survey_fiber}, 
are used. The central value of the overlap area
is used as the position where the particle passed the MD. 
The difference between track coordinates reconstructed in the fibre detector and the positions 
predicted by the EUDET telescope is used to determine the spatial resolution. 
Since the contribution of EUDET to the convoluted resolution is negligible,
these plots quantify the resolution of the MD.

Various effects degrade the theoretical resolution noticed in section~\ref{sec:main_detectors}.
The most relevant contributions are:
\begin{itemize}
  \item The measured fibre positions (section~\ref{sec:survey_fiber})
  deviate from their design values, sometimes up to a few 100~$\mu$m.
  This deteriorates the layer staggering which is a condition for the ideal resolution. 
  \item The fibre layer efficiency (section~\ref{sec:Efficiency}) is not 100\% due to the 
  non-scintillating cladding around the fibre core and the amplitude cut for the digitisation     
  of the fibre signal.
  \item The cross-talk reduces the primary fibre signal and generates hits at outlying
  positions (section~\ref{sec:cross_talk}).
  \item Physics processes like Coulomb scattering or generation of $\delta$-electrons affect the 
  particle trajectory or generate signals in the neighbouring fibres.
\end{itemize} 
The measured spatial resolution of all MDs ranges between 30~$\mu$m and 40~$\mu$m.           

\subsection{Edge sensitivity and distance measurement}
\label{sec:Edge}
\begin{figure}[h!]
  \centering
  \includegraphics[width=120mm]{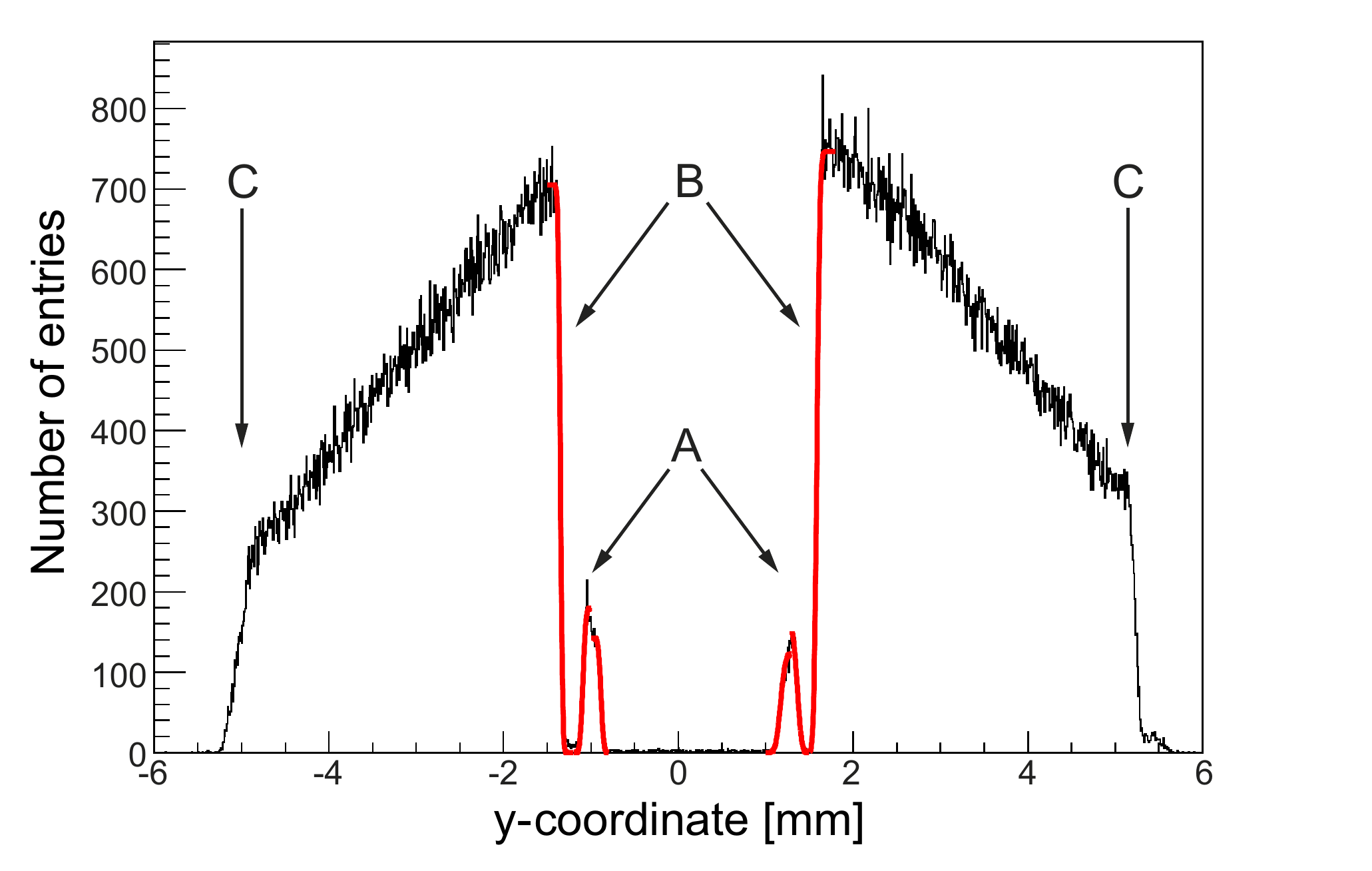}
  \caption{A vertical tomographic cross section of a  RP station with upper and lower MDs.
   Clearly visible are the RP bottom windows (A) and the lower MD edges (B). 
   At values of $\pm$~5~mm (C) the histogram is cut by the size of the EUDET trigger counters.}
   \label{fig:TB2010_RP-tomography}
\end{figure}

The reconstruction of the incident beam particle trajectories by the EUDET telescope 
allows a tomography of the fibre detectors inserted in the RPs.
The trigger condition requires a coincidence between the EUDET trigger counters, which cover 
the size of the pixel sensors, and the trigger counters of the MD fibre detectors described in 
section~\ref{sec:trigger_counters}. 
The vertical projection of the beam trajectories with a corresponding trigger signal 
is shown in figure~\ref{fig:TB2010_RP-tomography}.
In most cases the beam particle passes the EUDET telescope and the fibre detectors without 
interactions. The related track pattern indicates the lower (upper) edge of the upper (lower)
MD trigger counter and the outer edges of the EUDET trigger counters.  
If the beam hits the RP bottom windows interactions generate shower particles which are 
scattered into the upper or lower MD trigger counters. The track pattern of
these events shows the position of the upper and lower RP bottom windows as indicated 
in figure~\ref{fig:TB2010_RP-tomography}. 

The tomography pattern allows to measure positions inside the RP which are not accessible by
other measuring equipment.
For this purpose the edges of the pattern were fit with smeared step functions defined 
individually in the relevant ranges where the hit counts rise steeply.
The width of the smearing is about 20~$\mu$m and results from small differences of the size, 
the roughness and the reflectivity of individual fibre modules after the diamond cutting. 
Some discontinuities visible at the positions of the windows are resulting from the beam profile 
and the slight convex shape of the RP windows, discussed in section~(\ref{sec:pots_surveys}). 
The essential values from the fit are the positions of the MD edges and their distances 
to the RP bottom windows. For all combination of MDs and RPs the distance varies from 
150~$\mu$m to 250~$\mu$m. 
The fit confirmed the window thickness of 200~$\mu$m obtained by machining of the RP bottom 
surface. 
The distance between upper and lower MD edges changes with the RP movement. For the case shown 
in figure~\ref{fig:TB2010_RP-tomography} the distance between the upper and lower RP outer 
window faces was set to about 2~mm. 
     
Another important task of the test beam campaign was the calibration of the OD fibre 
positions with respect to the MD coordinate system.
For this purpose the MD coordinate system was aligned to the EUDET reference system. 
Then the OD fibre positions were scanned by EUDET tracks and compared with 
the values measured by microscope, described in section~\ref{sec:survey_fiber}.
The differences are in the order of 100~$\mu$m, and vary among all 
ODs. Applying corresponding corrections, the OD distance measurement 
differs less than 10~$\mu$m from the direct values given by the edges of upper 
and lower MDs.

%% file: sections/Experimental_setup.tex
\section{Experimental setup}
\label{sec:exp}
\begin{figure}[ht]
  \centering
  \includegraphics[angle=0,scale=0.12]{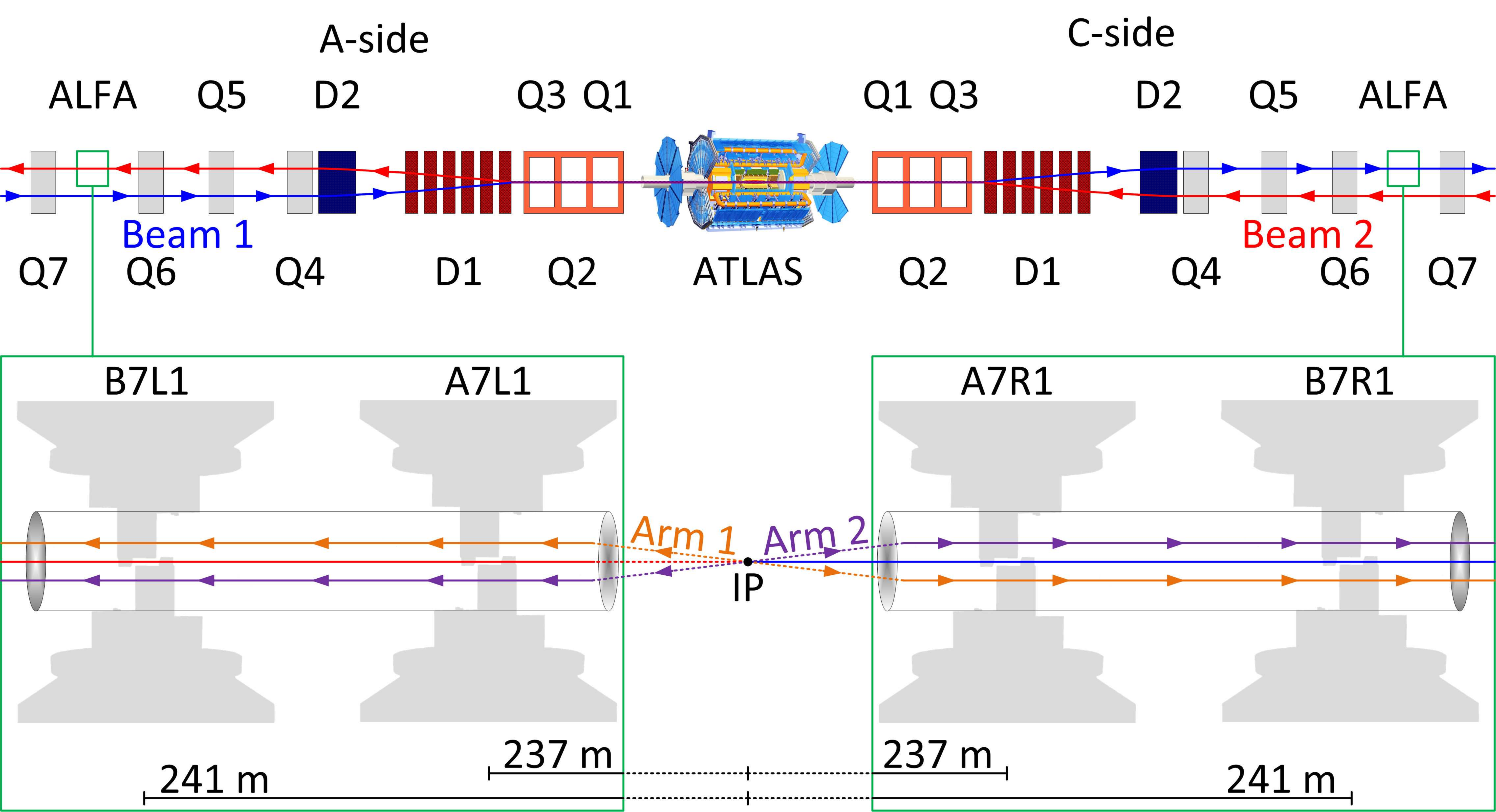} 
  \caption{A schematic view of the LHC beam lines up to the ALFA stations with the relevant magnets.
   The lower part indicates the station names and their positions in Run 1. Copyright CERN:
   reproduction of this figure is allowed as specified in the CC-BY-4.0 license.} 
  \label{fig:station_position}
\end{figure}

The ALFA system consists of four stations each equipped with two detectors in an upper and lower RP.
Two stations are placed in a distance of about 240~m at both sides
of the ATLAS interaction point, in the long straight LHC sections.
The distance between two stations was approximately 4~m during Run 1; it has been
increased to 8~m for Run 2 to improve the angular resolution.
The positions as well as the naming scheme of the stations are indicated in 
figure~\ref{fig:station_position}.
This chapter is split into three sections describing the main station components, the movement 
system and the calibration of the movement system.

\subsection{Roman Pot stations} 
\label{sec:roman_pot_stations}
The stations have been designed to satisfy the strict requirements of positioning precision 
and all aspects of the LHC operation, complying with ultra-high vacuum, 
impedance, temperature, electricity and safety constraints. 
The detectors inside the RPs and the readout electronics are physically separated from the LHC vacuum.

Each station is composed of a stiff main body and two RP flanges which can be moved vertically by 
step motors, described in section~\ref{sec:positioning_system}.  
\begin{figure}[ht]
  \centering
  \includegraphics[angle=0,scale=0.60]{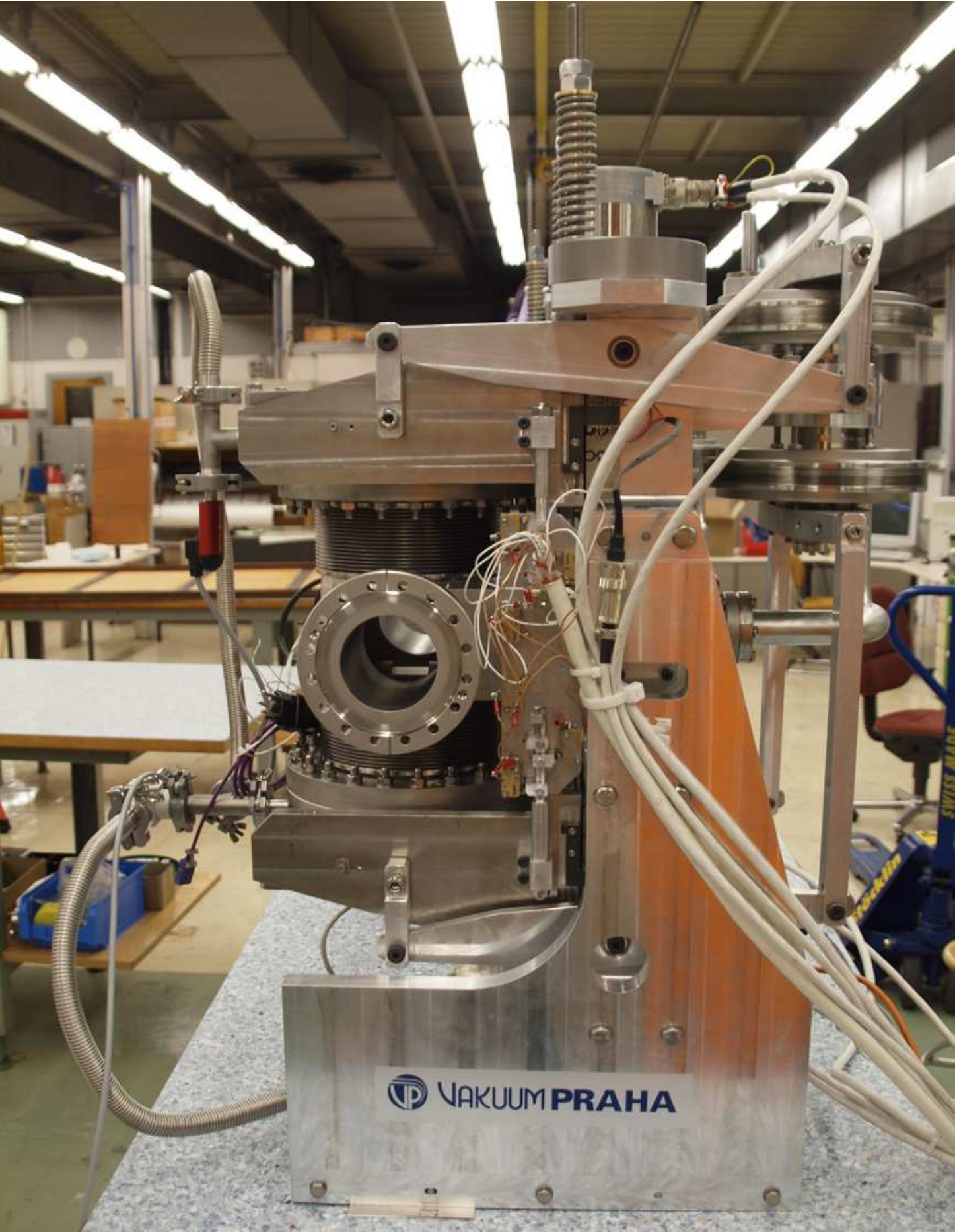} 
   \caption{A side view of a RP station in the lab. On the left side of the station body the round 
   flange of 
   the beam pipe, the upper and lower RP bellows and the RP flanges are visible.
   A lever arm with a pivotal point at the upper part of the station body 
   connects the RP flanges with the squeezed compensation bellows at the upper backside
   of the station. On top of the station two springs are mounted which support the 
   auto-retraction. In between the springs the step motor of the upper RP is installed.
   On the right side of the beam pipe flange the plate with the micro-switches is fixed.}      
  \label{fig:station_in_lab}
\end{figure}
The upper and lower flanges, which carry the RPs, are connected by flexible bellows to the 
beam pipe. 
The bellows are connected to the LHC vacuum chamber and allow an excursion of 50~mm towards the 
beam. A station in the assembling hall is shown in figure~\ref{fig:station_in_lab}.

Each station is equipped with a compensation system to counterbalance the 
atmospheric pressure on the RP and to ensure auto-retraction.  
It consists of two additional bellows placed at the backside of the station as shown in
figure~\ref{fig:station_in_lab}.
The compensation bellows are connected by lever arms 
to the RP bellows on the other side and balance the atmospheric pressure.  
Due to the larger size of the compensation bellows, the force on the RP bellows which pulls them 
with 2.7~kN  towards the beam is over-compensated by a outwards force of 3.6~kN.   
This ensures an auto-retractile movement in case of power loss.
The first station B7L1 was installed in the winter shutdown 2009 while the other three 
followed one year later.

\subsection{Positioning system}
\label{sec:positioning_system}
\begin{figure}[ht]
  \centering
  \includegraphics[width=140mm]{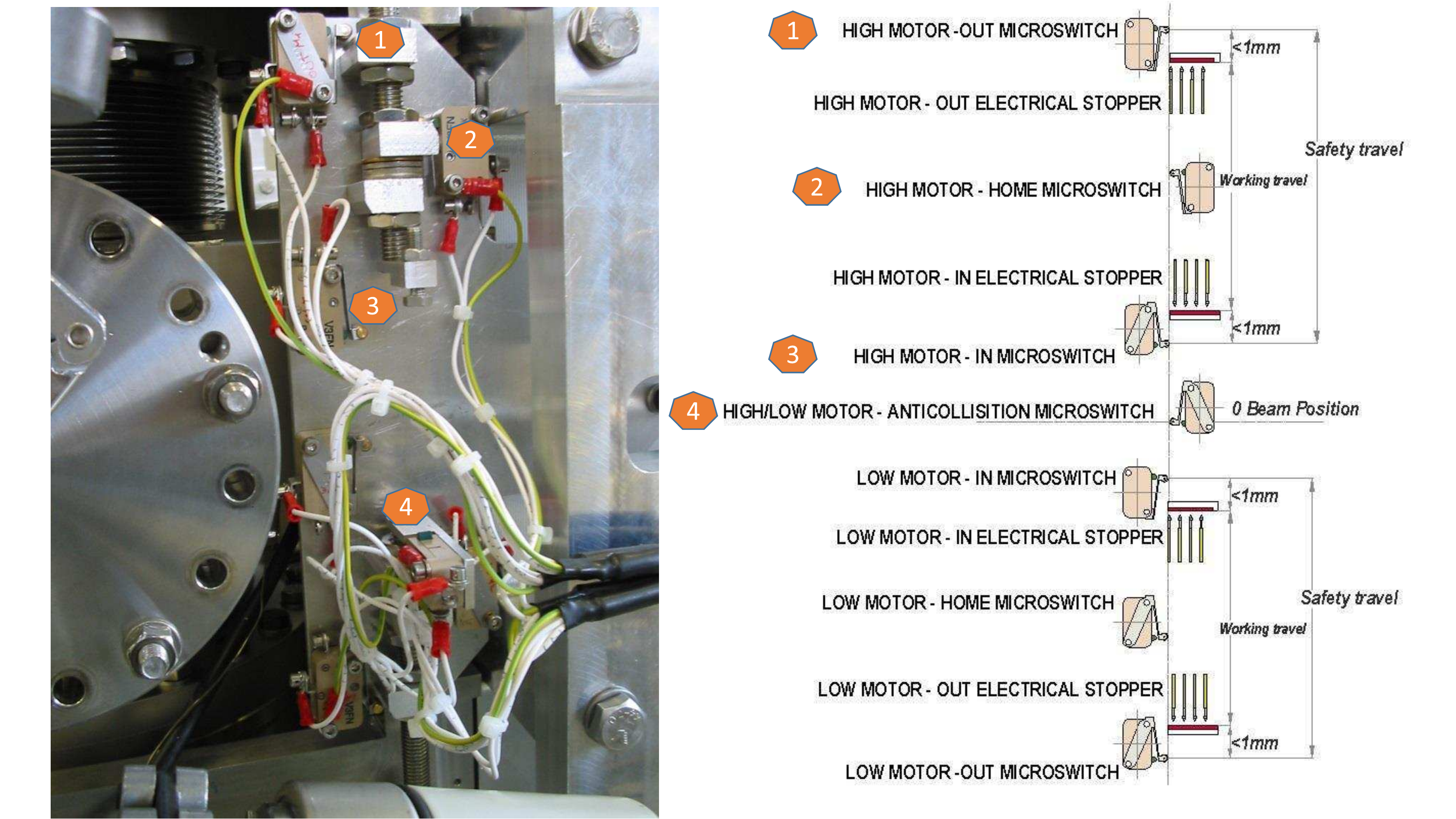} 
  \caption{The picture on the left side shows the practical implementation of the micro-switches
  attached to the RP station.
  The positions can be adjusted by means of screws and washers. On the right side the schematic 
  layout of all components is shown.} 
  \label{fig:switches}
\end{figure}

The station allows an independent movement of the upper and lower RPs via a high 
precision roller screw moved by a step motor.
The main components of the positioning system are the radiation hard motors, the rotary position 
sensors mounted on the motors, the Linear Variable Differential Transformers (LVDTs), and the 
switches and stoppers to control the movement range. 
Some details are listed in the following bullets:
\begin{itemize}
  \item The motors are expected to cope with an integrated radiation dose of 30~MGy. In operation, 
   one revolution corresponds to 400 motor steps, 5~$\mu$m each. 
   The expected lifetime of the motors is 15 million revolutions.
  \item Contactless resolvers have been selected as rotary position sensors due to their capability 
   to withstand high radiation doses. Although the resolvers were
   installed right from the beginning, they were operational only in Run 2.
  \item The LVDTs are used to measure the RP position with respect to the centre of the beam
   pipe. The position reproducibility is about 10~$\mu$m in periods of a few days when the data 
   runs are performed. In the long term larger drifts were observed. 
   Due to the high radiation environment, the readout electronics is placed outside the tunnel.
  \item The movement range of each RP is defined by an inner and outer electrical stopper.  
   The outer stopper serves as reference to follow the stability of the position values.
   To protect the motors against movement beyond the hard end-points, 
   micro-switches are placed in millimeter distance after the stoppers.  
   Accidental collision of the upper and lower RPs are blocked by an anti-collision switch.
   The system of micro-switches and stoppers is shown in figure~\ref{fig:switches}. 
\end{itemize}

\subsection{Calibration of the positioning system}
\label{sec:survey_station}
After the installation of the stations in the LHC tunnel the RP movement system was calibrated 
by a laser survey.\footnote{Leica LDT500, Leica Geosystems Inc., Norcross, US.} 
First, the coordinate system was defined by measuring the outer contours of the RP station.
Next, moving the RPs inwards the LVDT current ratio was recorded in steps of 2~mm (1~mm close 
to the beam centre) along the movement range. 
To include angular effects, the RP positions 
are derived from three laser targets which are fixed to the RP base plates. 
The position of these targets were measured in the laboratory by a 
3D device.\footnote{TESA Micro-hite 3D CMM.}
Combining the 3D survey of the targets with laser survey in the tunnel the LVDT current ratio 
is used to determine the absolute position of the RP bottom window with respect to the centre of 
the beam pipe. To get rid of fluctuations, typically in the order of 10-20$\mu$m, the 
RP positions as a function of the LVDT current ratio was parametrised by a third order polynomial.
As the LVDTs, also motor steps and resolvers were calibrated by the laser survey. 

The largest uncertainty on the RP positions results from the limited visibility of the contours 
defining the station coordinate system. 
Including the contributions from the 3D surveys of the RPs and the laser targets, 
it is in the order of 100~$\mu$m.     
The position recording from the LVDTs, the motor steps and the resolvers are used for the 
beam-based alignment before data taking, described in section~\ref{sec:BBA}.

%% file: sections/DCS.tex
\section{Detector Control System}
\label{sec:DCS}
The Detector Control System (DCS) is responsible to control  and to monitor all relevant
detector parameters using a Finite State Machine (FSM), to perform the alert handling and the data
archiving.
The local ALFA DCS server with a SLC6 Linux operation system is connected to the global ATLAS DCS 
which handles all ATLAS sub-detectors~\cite{bibDCS1}.

The front-end electronics is configured by sending appropriate parameter values to the 
FPGAs of the PMFs.
Parameters like gains, thresholds, control bits, trigger patterns, rates, latency 
have a direct impact on the data quality.
Other detector parameters like currents, voltages, temperatures, RP positions, etc. need to be 
followed for hardware failures. 
If they are outside a normal range an alarm is triggered and propagated through the FSM
alerting the operation team to take measures that assure the data quality.

Some detector components are directly handled through the Linux server, others which are
accessible only by a Windows interface are connected to a virtual machine with a Windows 
operations system.  
The ELMBs~\cite{bibElectronics4} for the configuration of the detectors, the temperature and 
radiation monitoring, 
the control of the VME crate with the  DAQ related front-end hardware and the controller 
of the low voltage power supply are directly connected by CAN-USB interfaces to the Linux server.
The virtual Windows machine controls through CAN bus connections the high-voltage power supplies, 
and through Ethernet connections the RP vacuum pumps and the fans for the air cooling.
The RP movement control is integrated in the LHC system and performed by two specific 
computers.\footnote{A National Instruments PXI which hosts the front-end real time application 
and a FESA server for the distributed information management.}

\begin{figure}[ht]
  \centering
  \includegraphics[width=150mm]{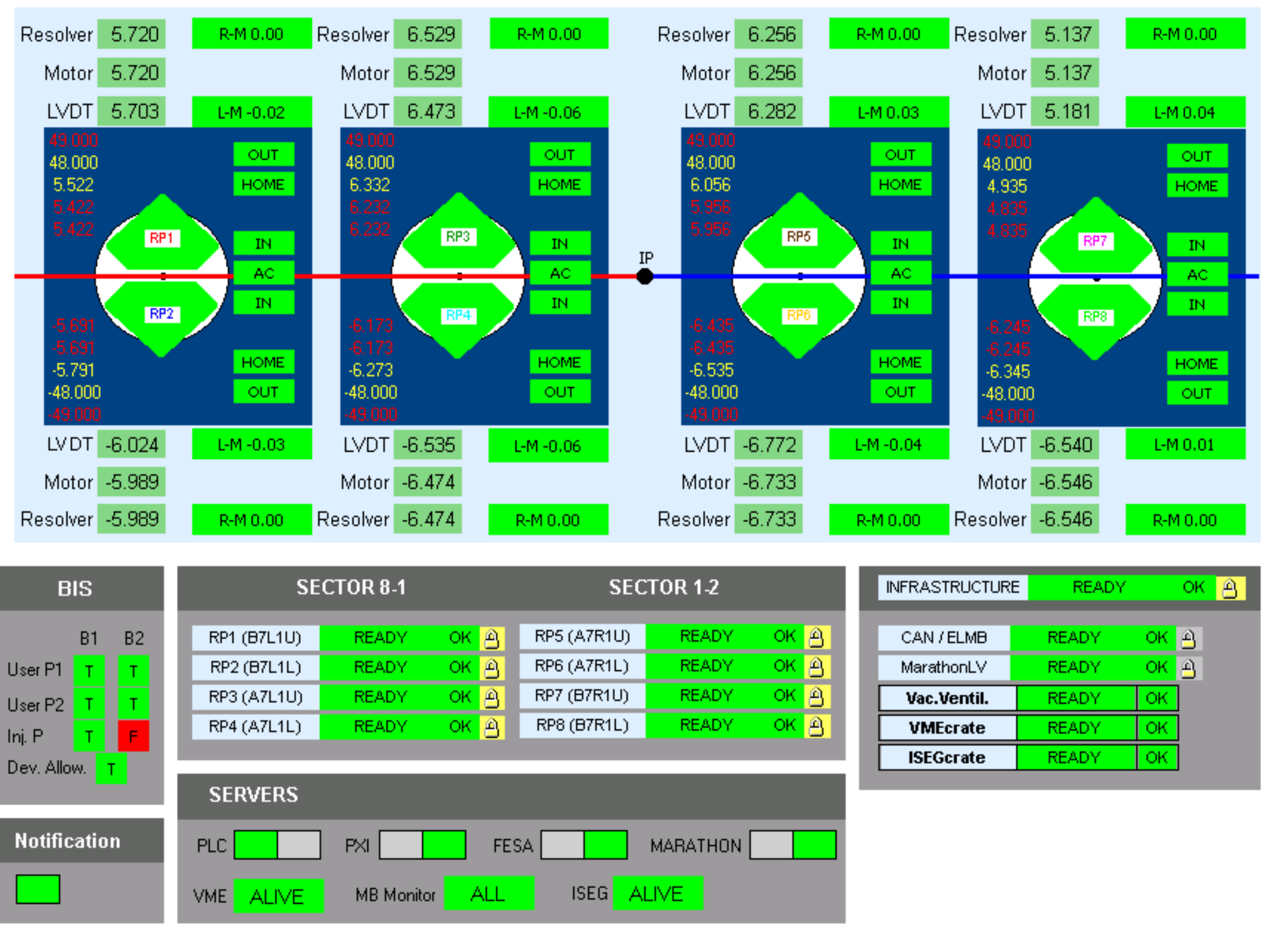}
  \caption{A cutout of the ALFA DCS top panel with the RP positions retrieved from the LVDTs, the 
   motor steps and the resolvers. The lower part indicates the status of the Beam Interlock System 
   (BIS), the power status of all detectors in sectors 8-1 and 1-2, the status of the infrastructure 
   components and of all servers. The snapshot is from the $\beta^* \approx$ 90~m run in October 
   2015 with RP positions at 10 $\sigma$ of the beam envelope.} 
\label{fig:dcs_hw}
\end{figure}

A cutout of the ALFA DCS top panel is shown in figure~\ref{fig:dcs_hw}. Various sub-panels allow
to access individual detectors or infrastructure components to configure or retrieve parameters.

%% file: sections/Trigger_DAQ.tex
\section{Trigger and data acquisition systems}
\label{sec:Trigger}

The signals of all eight MD trigger counters, described in section~\ref{sec:trigger_counters}, 
are discriminated, amplified and shaped in the triggerboards, referred in 
section~\ref{sec:trigger_board},
and fed into the Central Trigger Processor (CTP). 
Different combination of the eight input signals are used to select 
physics or background processes. The main trigger for elastic events consists of coincident signals 
from the upper detectors at ATLAS side A and the lower detectors at side C, and vice versa. 
To select diffractive events coincidences of both upper or lower detectors either at side A or 
side C are are used.
In addition, various 
trigger items were defined combining signals from other ATLAS sub-detectors or higher level objects 
like jets in coincidence with ALFA trigger signals. 
For the data readout two main chains were used: one with the data of all ATLAS sub-detectors for 
physics analyses and another only with ALFA data for systematic studies. 
In the second chain the small event size of about 5~kB allows high rates with looser 
trigger conditions which are useful for detector performance studies.                    

Due to the long distance of about 240~m from the ATLAS counting room the trigger latency is a 
primary concern. The main contributions to the ALFA latency are the time of flight from the
interaction point to the ALFA stations and the signal transmission to the ATLAS CTP with 800~ns and 
1030~ns, respectively. To minimize the transmission time special air-core cables which transfer 
signals with 91~\% of the speed of light are 
used.\footnote{Andrew Corporation, AVA5-50, HELIAX Coaxial Cable.} 
The signal handling in the front-end electronics, the cables and the evaluation of the trigger
logic adds up to another 270~ns and brings the ALFA latency beyond the standard ATLAS latency. 
In the special high-$\beta^*$ runs all ATLAS sub-detectors adapt to the ALFA latency.
In the standard LHC low-$\beta^*$ runs ALFA does not participate and the latency change is not
needed. 

\subsection {Trigger system} 
The triggerboard receives the raw trigger signals from the two MD and the two OD scintillator
tiles of a detector.
Details of the signal handling in the triggerboards are documented in 
section~\ref{sec:trigger_board}.  

\begin{figure}[h]
  \centering
  \includegraphics[width=150mm]{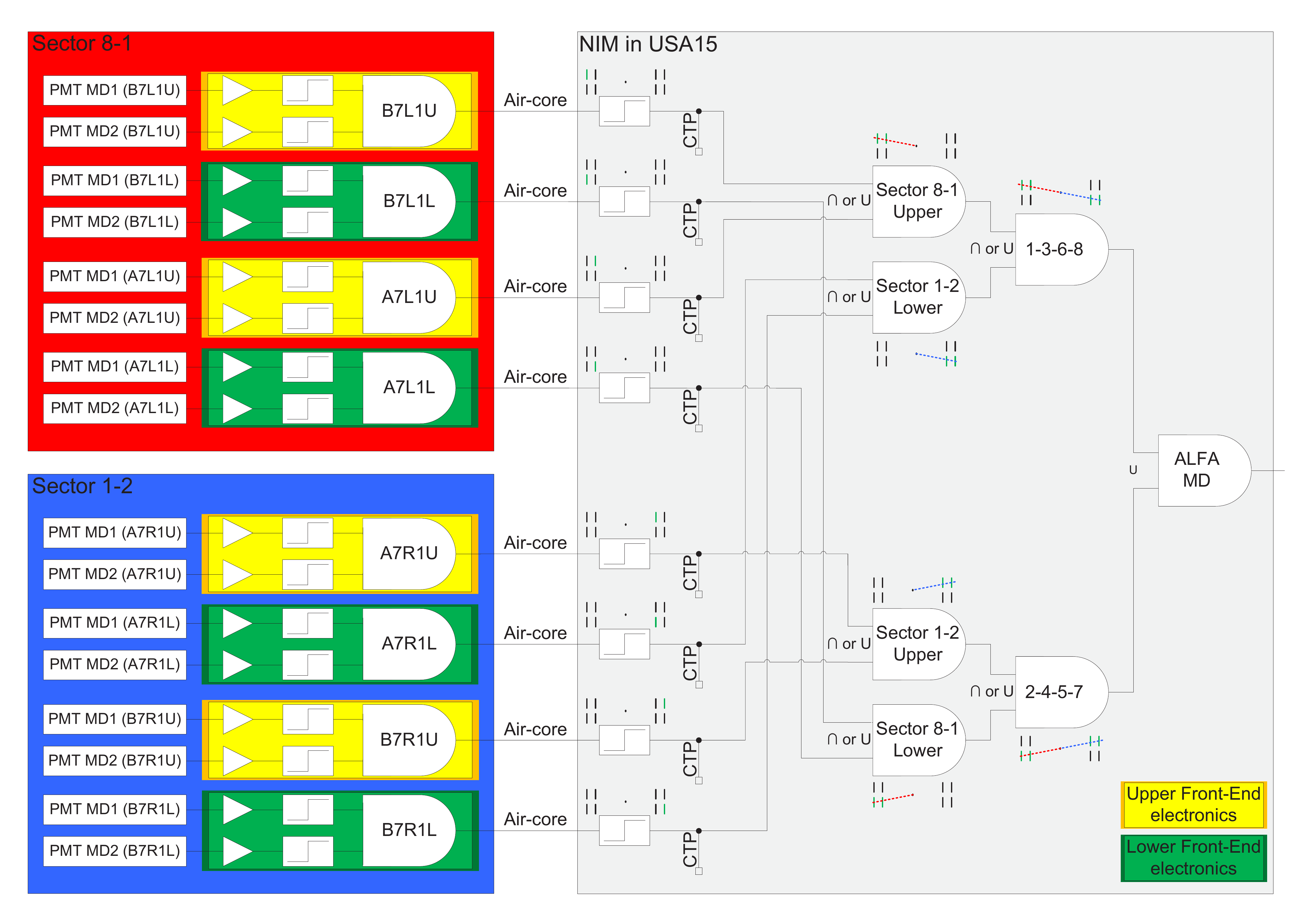}
  \caption {The logic scheme of the MD standalone trigger system. The two left boxes contain 
   the local signal handling in the triggerboards at the stations. 
   The right box contains the tree levels of coincidences for the formation of the final 
   trigger signal.}        
\label{fig:Trigger_MD}
\end{figure}

The scheme of the standalone trigger logic combining trigger signals from all MDs is shown 
in figure~\ref{fig:Trigger_MD}.
First a coincidence of the signals from the two MD trigger tiles is requested.
The resulting NIM signals are sent trough the 
air-core cables to the central counting room. Here they are split into a CTP and an ALFA  
standalone branch with three levels of coincidences as indicated in figure~\ref{fig:Trigger_MD}.
The standalone branch provides various options to trigger elastic events and is used for 
commissioning purposes. 
The CTP branch with the full spectrum of ATLAS trigger items is used for combined data taking 
with all ATLAS sub-detectors. 
The data taking for the distance measurement with the ODs, as described in 
section~\ref{sec:overlap_detectors}, was performed 
in standalone mode without the other ATLAS sub-detectors and has therefore
not to respect the L1 trigger latency. 
In 2011 conventional coaxial cables with a transit time of about 1200~ns were used  
for the signal transfer to the counting room.
Since 2012 the OD trigger signals are 
also transmitted by the fast air-core cables as the MD signals.
To distinguish MD and OD trigger signals different signal lengths were attributed to them.
The use of the fast air-core cables allows to perform 
the distance measurement in parallel with the physics data taking.

\subsection {Data acquisition system}
\label{sec:DAQ}
At each bunch crossing data from the scintillating fibres are fed into delay pipelines located in 
the PMFs of the front-end electronics.
There they are kept until a L1 trigger decision from the CTP arrives after a certain latency.
In case of a positive decision the data are transferred into the derandomising buffers of the 
motherboards, otherwise they are dropped. 
These buffers are the first part of the local DAQ system which provides data 
aggregation and transfer to the ATLAS-wide ReadOut System (ROS)~\cite{DAQ_1}

The data from the buffers are sent via optical links to the ReadOut Drivers (ROD)
which are the ATLAS-wide interface between the detector-specific infrastructure and the ROS 
computers.
ALFA utilizes the Muon ROD modules (MROD) which are designed for the ATLAS muon 
spectrometer~\cite{DAQ_2}.
The MROD module aggregates and frames the incoming data with the L1 
event and the bunch crossing numbers. 
In case of buffer overflow by reception of inconsistent data the module generates
a BUSY signal which can stop the ATLAS trigger generation to fix the source of corrupted data. 

The MROD module cooperates with another module - the TTC Interface Module (TIM) via a dedicated 
VME backplane. 
The TIM module receives L1 and bunch crossing information from the TTC system and broadcasts it 
to the MRODs.
To transfer the data from all detectors two MROD modules are used: one for the 
two upper detectors from the A-side and the two lower detectors from the C-side, 
and the other for the mirrored detector composition.
Such a connectivity allows to register elastic events within one module and facilitates 
the online monitoring.
Apart from the detector-specific MROD modules, the ALFA readout system utilises ATLAS-wide 
crate controllers and modules of the trigger and the TTC systems.

The software of the ALFA DAQ is residing on the ROD crate software framework 
and executed at the controller of the VME crate. It provides all functionality for the 
control of the MROD modules including configuration and communication at each transition of 
the DAQ system, i.e. transitions to the configure, start, or stop states of the DAQ. 
In addition, it allows to acquire data samples from the MRODs via the VME interface. 
This data can be used for the on-line monitoring of the data quality and give insight 
into the state of the DAQ system.

%% file: sections/Operation_experience.tex
\section{Operation experience and upgrades}
\label{sec:Operation}

After the installation and commissioning phases of the ALFA stations, data taking 
took place from 2011 to 2013 and was restarted with Run~2 in 2015.    
In the first part of this chapter we summarize the experience from alignment of the RPs and
the knowledge of the environmental conditions, like radiation levels and temperatures.  
The second part of the chapter is about the upgrades implemented during the long LHC break 
in the years 2013 and 2014. They include improvements of the heat protection, 
the angular resolution and the trigger performance.

\subsection{Beam-based alignment}
\label{sec:BBA}
A very precise positioning of the RPs with respect to the beams is mandatory for an optimal 
data taking and the safety of the LHC machine. 
This is the aim of the Beam-Based Alignment (BBA). 
For these procedure the laser-calibrated RP positions, described in 
Sec.~\ref{sec:survey_station}, have been used.
Later, for physics analysis, a track-based alignment is used to determine precisely the MD 
positions inside the RPs with respect to the beam~\cite{Elastics7}. 

In the BBA procedure the RPs are moved towards the beam until the outer RP windows scrape the 
edge of the collimated beam and induce a spiky signal in the Beam Loss Monitors (BLMs) 
downstream of the stations.
For the low intensity high-$\beta^*$ runs the BLM dump thresholds 
were increased by a factor 10 compared to standard running conditions. 
Close to the beam the movement step size is 10~$\mu$m resulting in a precision of about 
20~$\mu$m for the edge of the beam.
After the alignment of one station, the beams are reshaped by the vertical primary collimators
and the procedure is repeated for the remaining stations. 

The BBA procedure is done in a dedicated fill before the data taking with same beam conditions
used for the high-$\beta^*$ runs. 
From the vertical positions of the beam edges the centre and the width of the beam 
are obtained. These values are used for the positioning of the RPs in the subsequent fill for 
data taking.

\subsection{Radiation measurements}
\label{sec:radiation}
For the radiation measurement two different devices have been used;   
the Thermo-Luminescent Dosimeter (TLD) for integral dose measurement and the 
Radiation Monitor (RadMon) for on-line monitoring.

Two types of TLDs with different sensitivities were placed at positions compatible with the MDs 
in parking position: the TLD-100 and the TLD-800 with ranges from 10~$\mu$Gy to 10~Gy and 
0.5~mGy to 10$^5$~Gy, respectively.\footnote{GC Technology Messgeraete Vertriebs GmbH, Freidling 12,
D-84172 Buch am Erlbach.} 
In periods of about half a year the irradiated TLDs were replaced by fresh ones. After  
Run~1 the accumulated dose measured by the TLD-100 varied between 10~Gy and 30~Gy 
depending on the detector position.
\footnote{The TLD dose values are given in units Sv. To convert into units Gy a factor 
2 has been used, assuming mainly electro-magnetic radiation with an admixture of hadrons.}
Due to interactions at the inner stations the dose in the outer stations was 
about 10~\% to 20~\% higher. 
More striking is the 50~\% dose excess in the upper  
detectors compared to the lower ones due to the asymmetry of the beam background. 
The values from the TLD-800 are consistent with the TLD-100 results but are less reliable at 
such low doses.

The RadMons were installed at various positions around the stations A7L1 and B7L1:
two devices close to the detectors (comparable with the TLD positions), one between the beam 
pipes and three close to the motherboards and the PMFs.
They are equipped with two Radiation-sensing Field-Effect Transistors (RadFET) of 
different sensitivity.
In Run~1 only the high-sensitivity 
RadFETs were used~\footnote{LAAS 1600, Laboratory for Analysis and Architecture of Systems, 
Toulouse, France.}, while in Run~2 at positions with doses above 5~Gy the readout was switched 
over to the low-sensitivity RadFETs.\footnote{REM TOT501C type K, REM Oxford Ltd., United Kingdom.}
The dose values of the RadMons at detector positions confirmed the TLD values with an 
uncertainty of a few Gy.
The RadMons close to the front-end electronics indicated smaller doses at the level of a few Gy. 

The radiation hardness of scintillator material is in the range of a few kGy. Dedicated 
studies made with the fibres used for the ALFA trackers have shown a 10 \% efficiency loss
at 3~kGy. Hence the dose accumulated in Run~1 is 
far from a degrading impact on the tracking detectors. 

The part of the ALFA detector most sensitive to radiation is the front-end electronics. 
It is not designed as radiation-hard and sensitive components can start malfunctioning 
at doses in the order of 10~Gy. 

\subsection{Impedance measurements}
\label{sec:RF}
The installation of the RP stations impacts the LHC impedance budget.
In order to avoid the generation of beam instabilities and power deposition to the RPs, 
heating up the detector, laboratory tests have been performed.
A thin wire was stretched through the main chamber of the station 
and connected to the two ports of a vector network analyser which was used as RF source and 
signal analyser. 
This allows to determine the longitudinal and 
transverse impedance which couple to the LHC beam spectrum. 
By comparison to a reference signal 
from a bare beam pipe without the RP caverns the impedance can be calculated. 
Measurements at various RP positions with different conditions were performed: with the bare RP 
and with a semi-circular ferrite tile mounted to the RP bottom.  

The measurements without ferrite tiles have shown a number of resonances at frequencies 
between 0.6~GHz and 2.5~GHz.
Resonances at low frequencies between 0.6~GHz and 1~GHz are dangerous because they 
couple strongly to the beam harmonics.  
The installation of the ferrites reduced the resonances up to 1~GHz significantly. 
Consequently all RPs were equipped with ferrite tiles, as described in 
section~\ref{sec:pots_surveys}.

The RF losses due to the enhanced impedance results in a temperature increase in the 
material of the stations. 
The evolution of the temperature during the LHC fills is described in the next section. 

\subsection{Temperature measurements}
\label{sec:temperature}
All stations are equipped with temperature sensors at various points of interest, in particular 
at the front-end electronics, the station body, the RP walls and the titanium substrates of the 
fibre modules.\footnote{PT100 platinum resistance thermometers.} 
The sensors are connected to ELMBs and data transferred through CAN bus lines to the central DCS
for monitoring.

\begin{figure}[ht]
  \centering
  \includegraphics[angle=90,scale=0.58]{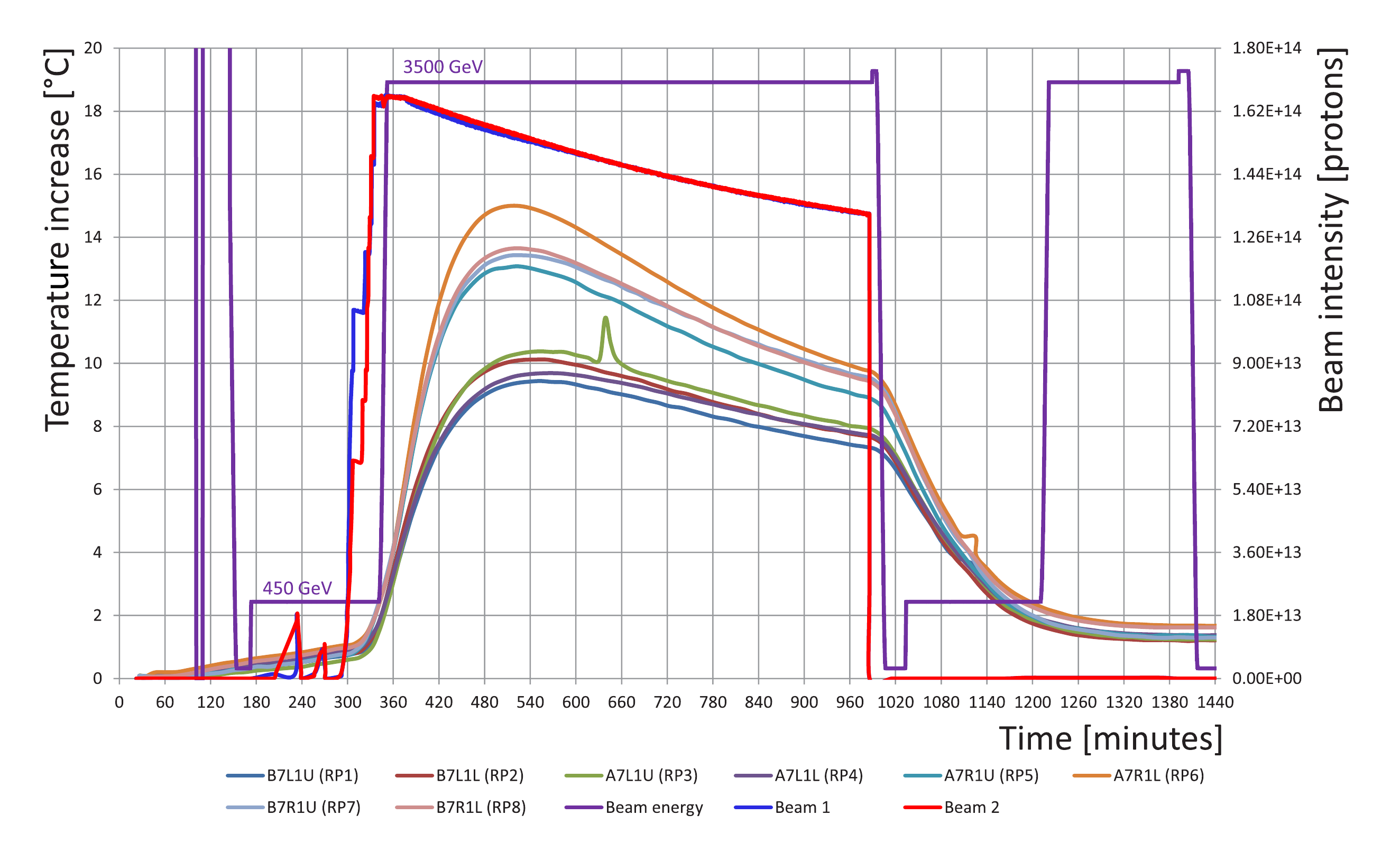}    
  \caption{The temperature evolution on the titanium substrates of all RP detectors during a 
   typical fill in Run~1. The individual detectors are labelled by different colour lines. 
   In the peak region the temperature of detectors on the C-side (RP5-RP8) is systematically a few 
   degrees higher compared to the detectors on the A-side (RP1-RP4). 
   The blue and red lines stand for the beam intensities, the purple line for the beam energy.}
  \label{fig:RP_temp_run1}
\end{figure}

Figure~\ref{fig:RP_temp_run1} shows the temperature increase at the titanium substrates of 
the MDs as a function of the time. 
Starting at injection of the beam the temperature increase reaches a maximum typically 3 hours 
after the energy ramp. As expected, the increase is higher at larger beam intensities but depends 
also on the details of the beam structure. 
During the standard LHC fills with high beam intensities in Run~1 when the RPs are in garage 
position, a dangerous temperature increase up to 20$^\circ$ was observed at the titanium substrates 
where the fibres are glued.
Adding the ambient tunnel temperature, at some places up to 25$^{\circ}$C, the detectors were 
heated up to 45$^{\circ}$C. 
The titanium substrates and the fibres glued on them have different thermal expansion coefficients
and the resulting stress can destroy the assembly. Laboratory tests have shown that the 
destructive phase begins at temperatures of about 45$^{\circ}$C. 
The temperature increase was traced back to the RF losses in the cavity housing the RPs with 
the detectors. Simulation indicated a power deposit of 10~W, 
which translates into the maximum observed increase of 20$^\circ$. 
The modifications to reduce the detector heating are described in the next section. 

\subsection{Upgrades for Run~2}
\label{sec:Upgrade}
The main effort in the break after Run~1 was dedicated to the heat protection. 
Extrapolations to Run~2 indicated a power deposit of up to 80~W in case of extreme beam 
conditions.
Other modifications were the displacement of the two outer stations and the improvement of 
trigger capabilities. 

\begin{figure}[ht]
  \centering
  \includegraphics[width=150mm]{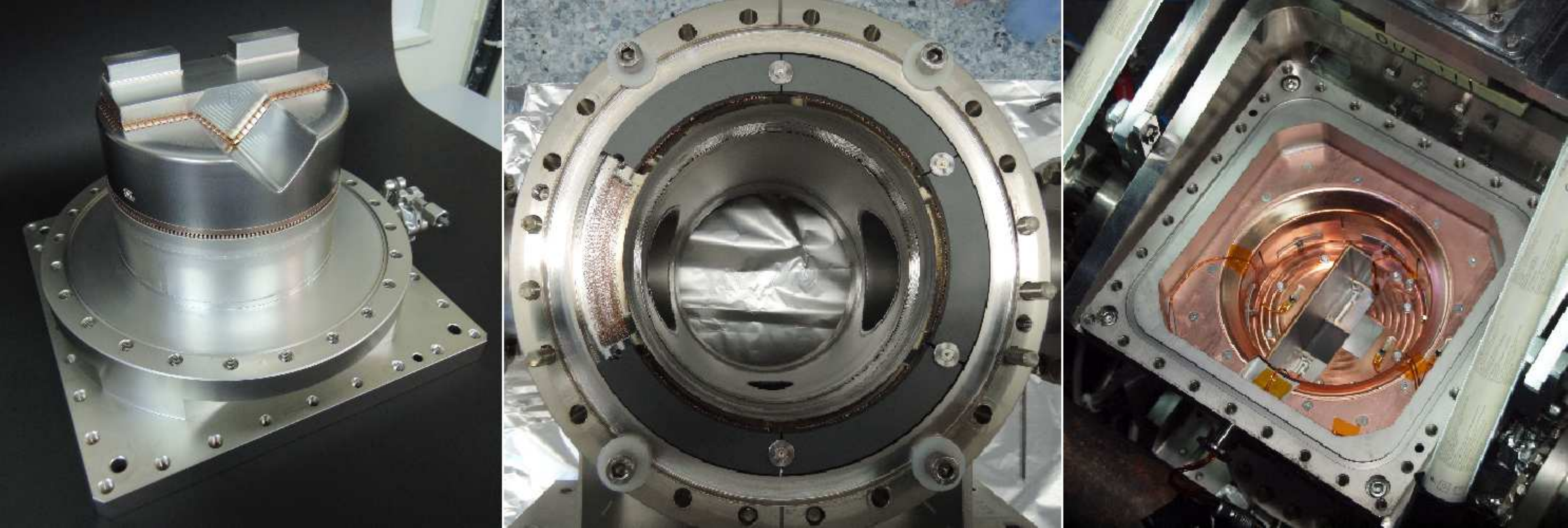} 
  \caption{Three modifications to improve the heat protection.
   From left to right: the RP-filler attached to the bare RP,
   the ring of ferrites distributed on the RP station flange
   and the copper heat distribution system with attached temperature probes.}
  \label{fig:alfa_upgrade_new}
\end{figure}

\begin{figure}[hb!]
  \centering
  \includegraphics[width=120mm]{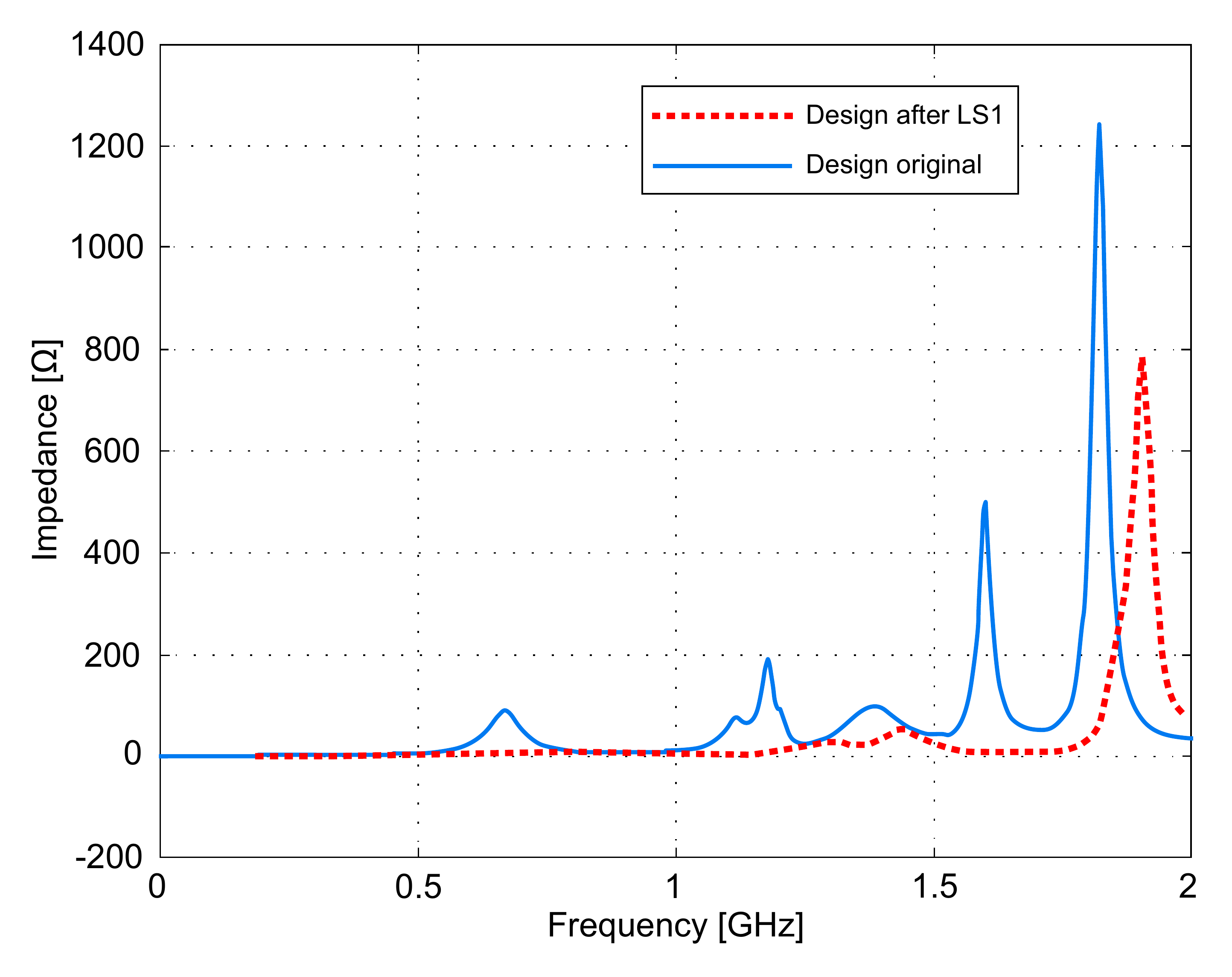} 
  \caption{The longitudinal impedance as a function of the frequency indicating the coupling 
   strength to the LHC beam spectrum. The blue curve describes the initial RP design, the red 
   curve stands for the modified design with a RP-filler.}
  \label{fig:alfa_frequency}
\end{figure}

To minimize the RF impact on the temperature four measures were performed.
The most important was to reduce the cavity volume and thereby the RF power deposit 
by extending the RP by a so-called RP-filler. 
In addition, the ferrites were mounted at positions further away from the detectors 
where the wake fields of the beams are absorbed in a more efficient way
\footnote{Nickel spinel TT2-111, specification http://www.trans-techinc.com, Adamstown, MD21710, US.}. 
Moreover, a heat distribution system made of thin copper plates was
installed inside the RP and the air cooling was intensified by additional fans 
and a better streaming to the RP flanges.
The three first modifications are illustrated in figure~\ref{fig:alfa_upgrade_new}. 
The RP-filler is made of titanium and connected by conductive copper-beryllium springs 
for proper grounding to the original 
RP.\footnote{Feuerherdt GmbH, D-12277 Berlin, Germany, http://feuerherdt.de.}  
The impact of the RP-filler on the resonance lines of the impedance spectrum is shown in 
figure~\ref{fig:alfa_frequency}. All resonance lines between 0.6~GHz and 1.8~GHz are 
strongly reduced. In particular those below 1~GHz, which have the strongest coupling 
to the beam spectrum and are most dangerous for heating, disappeared completely.
The second modification is related to the ferrites with completely revised position and shape. 
The semi-circular tile attached 
to the RP bottom has been replaced by a ring of six individual pieces located on the flange, 
visible in the central section of figure~\ref{fig:alfa_upgrade_new}.
At the new positions the heat is directly emitted to  
the stainless steel flanges where the air stream for cooling is circulating. 

After extensive lab tests, the stations were reinstalled in the LHC tunnel.
The temperature increase in a high intensity fill is below 5$^\circ$, far from the destructive
range. 
Another important effort for Run~2 was the displacement of the outer stations
B7L1 and B7R1 downstream from their original positions. In this way the distance between 
inner and outer station was enlarged from 4~m to about 8~m. 
Naturally, the angular resolution was improved by a factor two and 
results in a higher precision of the derived proton momentum transfer.  

The trigger upgrades concerned the dead time and the latency. 
The reduction of the dead time was achieved by upgrades in the firmware of the MAROC chip  
of the front-end electronics, described in Sec.~\ref{sec:Electronics}.
It was reduced from 550~ns to 88~ns. This allows triggering and data taking with a bunch 
spacing down to 100~ns.

The upgrade of the ATLAS CTP for Run~2 brought a latency penalty of about 75~ns what
increased the ALFA latency beyond the limits of the L1 trigger budget.    
For compensation a new back-end triggerboard was produced. 
It allows to bypass the standard trigger input stages and injects signals directly to the 
CTP core~\cite{Sune_2010}.
In addition, it replaces the NIM electronics used in the past for the separation of MD and OD 
trigger signals and allows the monitoring of rates per bunch crossing.

%% file: sections/Summary.tex
\section{Summary}
\label{sec:Summary}

The construction of the ALFA system and its integration in the infrastructure
of the ATLAS experiment is described. 
Two RP stations are placed at each side of the ATLAS detector at distances of 
about 240~m.
Each station consists of an upper and a lower RP which can be moved towards the beam in vertical 
direction. 
The RPs are equipped with tracking detectors to measure forward going protons from elastic or 
diffractive processes.
      
The MDs consist of two times ten layers 
of square 0.5$\times$0.5~mm$^2$ scintillating fibres orthogonally arranged  
perpendicular to the beam.  
By staggering the individual fibre layers a spatial resolution of about 35~$\mu$m is achieved. 
Each layer of 64 fibres is read out by one MAPMT.
The typical light yield per fibre is between 4~PEs and 5~PEs what results in a 
layer efficiency above 90~\%.
For the alignment of the detectors the distance between 
upper and lower MDs is measured by special ODs which are attached to the MDs. 
For triggering the detectors are equipped with scintillator tiles. The sensitive area of the 
main detectors is covered by two 3~mm thick tiles while a single tile is used for the ODs. 

The front-end electronics of the fibre readout is directly placed on the back-side of 
each MAPMT. 
A dedicated triggerboard receives the signals from the trigger tiles and prepares the 
transmission to the CTP.
If a L1 signal returns, the digitised fibre data are transmitted to the motherboard. 
Here the event is built and transmitted through an optical link to the 
ATLAS DAQ system. 

The positions of the RPs and their reproducibility are an essential parameters 
for the physics analysis and the safety of the LHC machine. 
A precision motor allows the movement of the RPs in steps of 5~$\mu$m from the garage position 
towards the beam and vice versa.
For the calibration of the RP position a special laser survey was performed.  
Combining this survey with the 3D measurements of detectors and RPs, the absolute position 
is known with a precision of about 100~$\mu$m. 

Due to the far position of the detectors the ALFA trigger signals arrive beyond 
the standard ATLAS latency budget. Fast air-core cables were installed to keep the ALFA latency 
within the maximum length of the buffering pipeline.
Since 2012 also the overlap trigger signals were sent trough the air-core cables to CTP, coded by 
a different signal length.  

To ensure the functionality of detectors and electronics over many years 
the control of radiation dose and temperature is vital. 
For dose measurements crystal-based TLDs and electronic RadMons for online monitoring have been 
used.  
In Run~1 both sensors indicate doses between 10~Gy and 30~Gy at the detector positions. 
The temperature was monitored by PT100 elements attached to all relevant parts of the station.
A critical temperature increase up to $45^\circ$ related to the RF 
losses at high beam intensities was observed with RPs in garage position.  
Protection against heating was 
achieved by multiple measures: a RP-filler to reduce the cavity volume, 
a modified ferrite system, a heat distribution system and an optimized 
air-stream for the cooling.  

The main task of ALFA is the measurement of the proton-proton total cross section at all LHC 
energies. The results of the Run~1 data taking at 7~TeV and 8~TeV are 
published~\cite{Elastics7},~\cite{Elastics8}.
In Run~2 the first data taking at 13~TeV was performed with a $\beta^* \approx$ 90~m optics.
Recently, in October 2016, data were taken with an ultra-high $\beta^* \approx$ 2500~m. 
Both data sets allow the measurement of the elastic and total cross sections. In addition, 
the ultra-high $\beta^*$ value gives access to the Coulomb region at very small values of the
proton momentum transfer. This opens the possibility to measure the 
luminosity solely based on elastic scattering.